\documentclass[aps,pra,reprint,twocolumn,superscriptaddress,floatfix,nofootinbib,longbibliography]{revtex4-2}
\usepackage{lmodern} 
\usepackage{float}
\usepackage{microtype}
\usepackage{graphicx, amsfonts, amsthm, xr}
\usepackage{array}
\usepackage[dvipsnames]{xcolor}
\usepackage{epsfig, amsmath, amssymb, upgreek}
\usepackage[english]{babel}
\usepackage[T1]{fontenc}
\usepackage{mathtools}
\usepackage{comment}
\usepackage{braket}
\usepackage{lipsum}
\usepackage{soul}
\usepackage{bm}

\usepackage[inline]{enumitem}
\setlist[enumerate]{label=(\roman*)}

\usepackage{tikz}
\usepackage{cancel}
\usepackage{siunitx}
\usepackage{multirow}
\usepackage{gensymb}
\usepackage{tabularx}
\usepackage{booktabs}
\usetikzlibrary{fadings}
\usetikzlibrary{patterns}
\usetikzlibrary{shadows.blur}
\usetikzlibrary{shapes}

\IfFileExists{stmaryrd.sty}{\usepackage{stmaryrd}}{}

\renewcommand{\ket}[1]{\left| #1 \right\rangle}
\renewcommand{\bra}[1]{\left\langle #1 \right|}

\DeclareMathOperator{\Tr}{Tr}  

\newcommand{\SU}[1]{\mathrm{SU}(#1)}
\newcommand{\U}[1]{\mathrm{U}(#1)}
\newcommand{\Z}[1]{\mathbb{Z}_{#1}}



\definecolor{boxback}{HTML}{FFF8B5}
\definecolor{applegreen}{rgb}{0, 0.5, 0.0}
\definecolor{smoothred}{HTML}{C5232F}
\definecolor{mygreen}{rgb}{0,0.5,0}
\definecolor{myblue}{rgb}{0,0,0.75}
\definecolor{mymagenta}{cmyk}{0,1,0,0.12}

\usepackage[bookmarks=true,colorlinks,linkcolor=OrangeRed,urlcolor=NavyBlue,citecolor=RoyalBlue]{hyperref}
\usepackage[capitalise]{cleveref}
\Crefname{figure}{Fig.}{Figs.}
\Crefname{section}{Sec.}{Secs.}
\Crefname{appendix}{App.}{App.}
\makeatletter
\AddToHook{cmd/appendix/before}{\def\cref@section@alias{appendix}}
\makeatother

\usepackage{orcidlink}
\newcommand{\orcidpietro}{\orcidlink{0000-0001-5279-7064}}
\newcommand{\orcidsimone}{\orcidlink{0000-0002-8882-2169}}
\newcommand{\orcidmarco}{\orcidlink{0000-0002-4544-3513}}
\newcommand{\orcidpeter}{\orcidlink{0009-0009-7020-7246}}
\newcommand{\orcidmattia}{\orcidlink{0009-0006-1796-6989}}

\definecolor{todogray}{gray}{0.4}
\definecolor{todored}{rgb}{0.6,0,0}

\usepackage{fontawesome5}

\newcommand{\letterimagetext}[1]{\textcolor[RGB]{236,22,91}{#1}}
\newcommand{\letterimagecaption}[1]{\textsf{\textbf{(#1)}}}

\begin{document}

\title{Preparation and detection of quasiparticles for quantum simulations of scattering}

\author{Mattia Morgavi\thinspace\orcidmattia}
\affiliation{Dipartimento di Fisica e Astronomia, Università degli Studi di Padova, Padova, Italy}
\affiliation{Istituto Nazionale di Fisica Nucleare (INFN), Sezione di Padova, I-35131 Padova, Italy}

\author{Peter Majcen\thinspace\orcidpeter}
\affiliation{Dipartimento di Fisica e Astronomia, Università degli Studi di Padova, Padova, Italy}
\affiliation{Istituto Nazionale di Fisica Nucleare (INFN), Sezione di Padova, I-35131 Padova, Italy}

\author{Marco Rigobello\thinspace\orcidmarco}
\affiliation{Max-Planck-Institut für Quantenoptik, Hans-Kopfermann-Str.~1, D-85748 Garching, Germany}

\author{Simone Montangero\thinspace\orcidsimone}
\affiliation{Dipartimento di Fisica e Astronomia, Università degli Studi di Padova, Padova, Italy}
\affiliation{Istituto Nazionale di Fisica Nucleare (INFN), Sezione di Padova, I-35131 Padova, Italy}
\affiliation{Padua Quantum Technologies Research Center, Università degli Studi di Padova, Italy}

\author{Pietro Silvi\thinspace\orcidpietro}
\affiliation{Dipartimento di Fisica e Astronomia, Università degli Studi di Padova, Padova, Italy}
\affiliation{Istituto Nazionale di Fisica Nucleare (INFN), Sezione di Padova, I-35131 Padova, Italy}
\affiliation{Padua Quantum Technologies Research Center, Università degli Studi di Padova, Italy}

\date{\today}

\begin{abstract}
    We introduce a method for the selective preparation and detection of quasiparticle wave packets, based on creation operators that generate dressed, localized excitations on top of interacting vacua of (quasi-)one-dimensional quantum lattice theories.
    This method exploits maximally localized Wannier functions (MLWFs) constructed from quasiparticle bands at intermediate system sizes, enabling the construction of unitary local dressed creation operators. The algorithm allows for species-resolved wave-packet preparation and detection, enabling the separation of known quasiparticle contributions from unknown resonances.
    We test this approach with matrix product states (MPS) on pure hardcore Hamiltonian QCD on a ladder lattice, detecting scattering outputs and mass resonances.
\end{abstract}

\maketitle

Predicting from first principles the real-time dynamics of quantum chromodynamics (QCD) --- including string breaking \cite{Bali2001QCDForcesHeavy,Greensite2011IntroductionConfinementProblem}, hadronization \cite{Metz2016PartonFragmentationFunctions,Albino2010HadronizationPartons} and thermalization processes \cite{bergesTurbulentThermalizationProcess2014,Banuls2020SimulatingLatticeGauge} --- remains a major challenge in high-energy physics; yet it is crucial for understanding the phenomenology of scattering events, such as heavy-ion collisions \cite{Busza2018HeavyIonCollisions,Gelis2010ColorGlassCondensate}.
Lattice Gauge Theory (LGT) 
\cite{Wilson1974ConfinementQuarks, Rothe2012LatticeGaugeTheories}
provides a powerful non-perturbative framework for 
QCD simulations.
Although conventional Monte Carlo approaches 
\cite{Creutz1980MonteCarloStudy,
Creutz1983MonteCarloComputations,
Metropolis1953EquationStateCalculations} 
are severely limited by the sign problem 
\cite{Loh1990SignProblemNumerical,
Troyer2005ComputationalComplexityFundamental, 
Barbour1986ProblemsFiniteDensity, 
Keldysh2024DiagramTechniqueNonequilibrium}
in out-of-equilibrium scenarios \cite{Polkovnikov2011ColloquiumNonequilibriumDynamics},
tensor-network (TN) 
\cite{Fannes1992FinitelyCorrelatedStates,
Ostlund1995ThermodynamicLimitDensity,
White1992DensityMatrixFormulation,
Abel2025RealTimeScatteringProcesses,
Schollwock2005DensitymatrixRenormalizationGroup,
Haegeman2011TimedependentVariationalPrinciple,
Haegeman2012VariationalMatrixProduct} 
and quantum-computing (QC) methods
\cite{Feynman2018SimulatingPhysicsComputers,
Lloyd1996UniversalQuantumSimulators,
Bauer2023QuantumSimulationFundamental} 
avoid this obstruction and have sparked renewed interest in 
Hamiltonian formulations of LGT
\cite{Kogut1975HamiltonianFormulationWilsons,
Kogut1979IntroductionLatticeGauge,
Chandrasekharan1997QuantumLinkModels,
Zohar2015FormulationLatticeGauge,
Buyens2017FiniteRepresentationApproximationLattice}.

Tensor networks have achieved state-of-the-art accuracy
for equilibrium properties of the massive Schwinger model \cite{Banuls2013MassSpectrumSchwinger,buyensMatrixProductStates2014,banulsChiralCondensateSchwinger2016},
and have been successfully applied also to higher dimensions
and dynamics, in both Abelian and non-Abelian LGTs
\cite{Felser2020TwoDimensionalQuantumLinkLattice,
Magnifico2021LatticeQuantumElectrodynamics,
Cataldi20232+1DSU2YangMills,
Rigobello2023Hadrons1plus1DHamiltonian}.
While real-time evolution remains challenging due to 
entanglement growth, the nature of this limitation 
is well understood 
\cite{Eisert2010ColloquiumAreaLaws,
vidalEfficientSimulationOneDimensional2004}.
Notably, in one-dimensional settings, real-time scattering 
dynamics typically does not generate rapid 
entanglement growth \cite{Rigobello2021EntanglementGeneration1+1D,
Belyansky2024HighEnergyCollisionQuarks}.
At the same time, quantum-simulation platforms offer a  
complementary route to real-time dynamics \cite{Byrnes2006SimulatingLatticeGauge,
zoharQuantumSimulationsGauge2013,
Bauer2023QuantumSimulationFundamental}, 
while TN methods remain crucial for cross-verification 
and benchmarking of results.

The controlled preparation of quasiparticle wave-packet input states above dressed vacua represents a key challenge for simulating real-time scattering phenomena \cite{Halimeh2025QuantumSimulationOutofequilibrium}.
Within the TN framework, the tangent-space quasiparticle ansatz allows for the construction of wave-packets on infinite matrix product state (MPS) backgrounds \cite{Haegeman2012VariationalMatrixProduct,Damme2021RealtimeScatteringInteracting,Milsted2022CollisionsFalseVacuumBubble,Belyansky2024HighEnergyCollisionQuarks,Jha2025RealTimeScatteringIsing, white2026sitebasisexcitationansatz}.
However, this method is not directly applicable to quantum simulation setups, motivating a recent development of more QC-friendly strategies, such as ad-hoc model-dependent algorithms 
\cite{Pichler2016RealTimeDynamicsU1,
Surace2021ScatteringMesonsQuantum,
Rigobello2021EntanglementGeneration1+1D,
Davoudi2024ScatteringWavePackets,
Papaefstathiou2025RealtimeScatteringLattice,
Davoudi2025QuantumComputationHadron,
Barata2025RealtimeSimulationJet,
Calliari2025QuantumSimulatingContinuum},
tunneling-breaking strategies 
\cite{Su2024ColdAtomParticleCollider,
Joshi2025ProbingHadronScattering,
Schuhmacher2025ObservationHadronScattering},
Givens rotations \cite{Chai2025FermionicWavePacket,Chai2025ScalableQuantumAlgorithm}, adiabatic switching \cite{Ingoldby2025RealTimeScatteringQuantum,Pavesic2025ScatteringInducedFalse} and W-states methods \cite{Farrell2025DigitalQuantumSimulations}.
With these approaches, scattering simulations have been implemented mostly in spin chains and (1+1)D Abelian LGTs (e.g., bosonized Schwinger model), in a $\SU{2}$ gauge theory \cite{Barata2025HadronicScattering1+1D}, in the two-dimensional Ising model 
\cite{Pavesic2025ScatteringInducedFalse}, and in continuum-limits of scalar field theories \cite{Jha2025RealTimeScatteringIsing,Calliari2025QuantumSimulatingContinuum}.
Despite recent progress, the development of model-independent approaches for quantum simulation platforms remains a significant open challenge \cite{Halimeh2025QuantumSimulationOutofequilibrium, turco2024quantumsimulationofbound, turco2025creationofwavepackets}.

\begin{figure}[t]
    \centering
    \hspace*{-5mm}\includegraphics{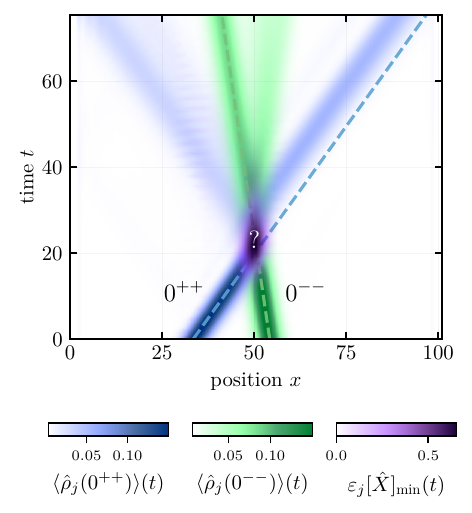}
    \caption{Real-time scattering of scalar-pseudoscalar hardcore $\SU{3}$ glueballs on ladder geometry (OBC) with $L=101$ plaquettes.
    MPS states are evolved via TDVP (max bond dimension $\chi = 100$, cutoff $\epsilon = 10^{-10}$). 
    The input state is prepared with the method described in \cref{sec:method},
    while single-quasiparticle densities $\langle \hat \rho_j \rangle$ and resonance detection are described in \cref{subsec:detection} and Appendix \ref{app:resonances}, respectively.
    The free 
    quasiparticle paths (dashed lines) are obtained by interpolation of the single-quasiparticle density before the collision.}
    \label{fig:face}
\end{figure}

In this work, we propose a novel method to efficiently approximate the action of quasiparticle creation operators on large system vacua by means of Wannier localization
\cite{wannierStructureElectronicExcitation1937,
Kohn1959AnalyticPropertiesBloch,
Marzari1997MaximallyLocalizedGeneralized,
Marzari2012MaximallyLocalizedWannier}
in systems accessible to exact diagonalization (ED) 
or Krylov methods.
These creation operators are applied at large-system sizes to (a) prepare wave-packets of desired momentum, position, and quasiparticle species; and (b) detect single-quasiparticle densities of specific quasiparticle species (see \cref{fig:face}), as well as projecting states into single- or multi-quasiparticle sectors.
The method is compatible with local gauge symmetries, different TN geometries (such as tree tensor networks \cite{Silvi2019TensorNetworksAnthology} and projected entangled pair states  \cite{Orus2014PracticalIntroductionTensor}) and is generalizable to higher dimensions.
Moreover, the output dressed creation operator is unitary and thus directly implementable in quantum circuits.
The algorithm is model independent and relies on two mild assumptions: (a) spatial homogeneity of the system and (b) spectral distinguishability of the single-quasiparticle bands of the input scattering quasiparticles.

Using MPS simulations, we test this approach in a non-Abelian quasi-(1+1)D LGT, corresponding to a pure $\SU{3}$ Yang-Mills theory on a ladder geometry and minimally truncated gauge field (i.e., hardcore-gluon truncation \cite{Rigobello2023Hadrons1plus1DHamiltonian,Cataldi20232+1DSU2YangMills}). 
As a benchmark, we also consider its Abelian $\Z{3}$ counterpart.
We identify the $\SU{3}$-glueball branches of the spectrum, and simulate the real-time scattering of glueballs, observing non-trivial interacting behavior in contrast to the $\Z{3}$ case in the weak coupling limit $g^2 \to 0$, suggesting hardcore glueballs interact even in the continuum limit.
We map the $\Z{3}$ and $\SU{3}$ ladder onto a one-dimensional chain, as previously done in 
\cite{Pradhan2024DiscreteAbelianLattice,hayataFloquetEvolution$q$deformed2025}.
This mapping restricts the system to the gauge-invariant sector, setting a concrete strategy for the implementation of this $\SU{3}$ model on a quantum simulator.
The algorithm is presented in a model-independent way.

The paper is organized as follows. \Cref{sec:method} presents the method: quasiparticle spectra from ED, Wannier localization, dressed creation MPO, and detection strategy. 
\Cref{sec:model} introduces the model and a duality mapping between the truncated $\SU{3}$ LGT on a ladder and a qutrit chain.
\Cref{sec:results} reports the results: glueball spectra, Wannier localization and creation operator construction, scattering simulations, and quasiparticle detection.
\Cref{sec:conclusions} summarizes this work and gives an outlook.
Technical details on the Hamiltonian formulation, tensor network methods, and the construction of the creation operator are collected in the appendices.

\section{The method}

\label{sec:method}

\begin{figure*}
    \centering
    \hspace*{-5mm}\includegraphics[scale=1.07]{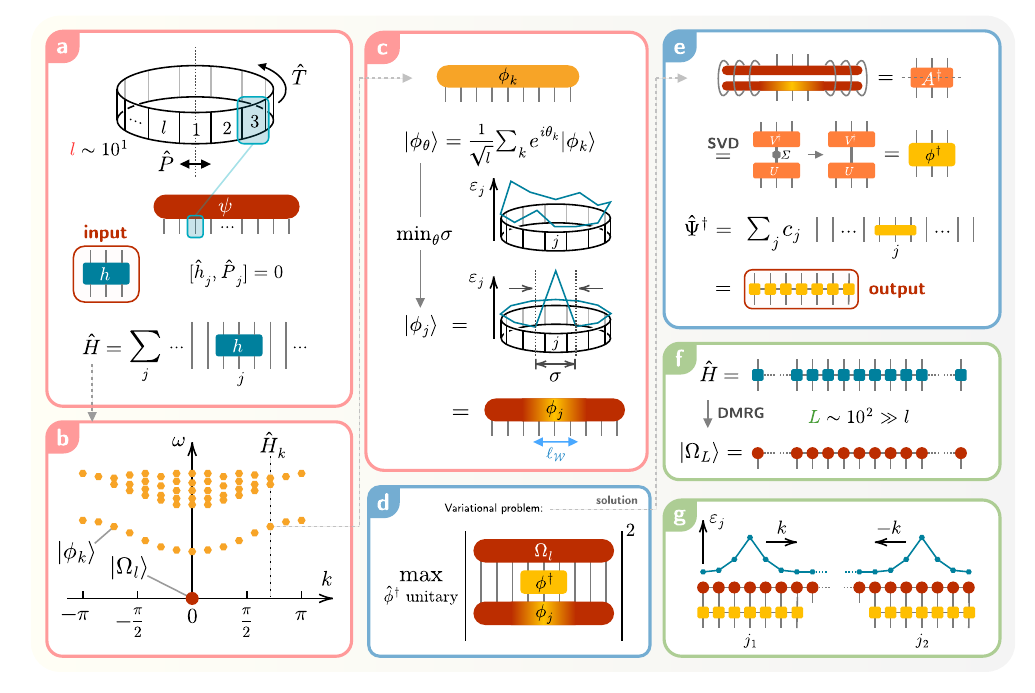}
    \caption{Schematic summary of the model-independent algorithm for generating the scattering input states. 
    \letterimagecaption{a} The states of a (quasi)-1D PBC quantum system of intermediate size $l$ are encoded in a vector representation. 
    The unique input is the local Hamiltonian $\hat{h}$, if possible, assumed to be parity invariant. 
    From $\hat{h}$, the translationally invariant Hamiltonian $\hat{H}$ is constructed. 
    \letterimagecaption{b} $\hat{H}$ is projected onto momentum sectors $\hat{H}_{k}$, and the spectrum $(\omega, k)$ at intermediate size is computed. 
    The vacuum $|\Omega_l \rangle$ and the Bloch states $|\phi_{k} \rangle$ of a single-quasiparticle band $\phi$ are selected.
    \letterimagecaption{c} Wannier states $|\phi_{\theta} \rangle$ are constructed from $|\phi_{k} \rangle$. 
    A spread functional $\sigma^{2}$ is variationally minimized to obtain MLWFs $|\phi_{j} \rangle $ around a site $j$. 
    \letterimagecaption{d} The variational problem is set: we look for a creation operator such that $\hat{\phi}^\dagger |\Omega_l \rangle =|\phi _{j} \rangle $ with maximum fidelity. 
    \letterimagecaption{e} To find such an operator, we perform the SVD of the partial trace $\hat{A}^\dagger$, as defined in the figure. 
    Thus, we set all the singular values to $1$; the result is the creation operator $\hat{\phi}^\dagger$.
    Finally, wave-packet creation operators $\hat{\Psi}^\dagger$ are constructed from $\hat{\phi}^\dagger$.
    \letterimagecaption{f} A large system of size $L\gg l$ is considered, finding the vacuum $|\Omega_L \rangle $ with DMRG.
    \letterimagecaption{g} The wave-packet creation operators are applied to $|\Omega_L \rangle $, producing the desired input scattering state.}
    \label{fig:algorithm}
\end{figure*}

Consider a quasi-one-dimensional lattice model with uniform local dimension $d$ and translational invariant Hamiltonian $\hat H$. The method consists in identifying quasiparticle excitations in a finite system, localizing them into Wannier states to define creation operators, and using these to construct scattering states in a larger system.
The main steps of the method can be summarized as follows:
\begin{enumerate}[label=(\Alph*)]
    \item solve with ED a system of intermediate size $l$ (the number of sites, length scale of red-contour panels in \cref{fig:algorithm}), in periodic boundary conditions (PBC, \cref{fig:algorithm}\letterimagetext{a}), finding the low-energy spectrum and selecting the quasiparticle excitation bands of interest (\cref{fig:algorithm}\letterimagetext{b});
    \item perform a maximal localization of the Wannier function (\cref{fig:algorithm}\letterimagetext{c}), corresponding to a selected band, by minimizing a spread functional; the maximally localized Wannier function must have a support $\ell_\mathcal{W} \ll l$ (length scale of blue-contour panels in \cref{fig:algorithm}) above the vacuum;
    \item solve a variational problem (see \cref{fig:algorithm}\letterimagetext{d}) for the extraction of a unitary creation operator from the Wannier state, which acts with a support $\ell_\mathcal{W}$, and construction of a wave-packet creation operator $\hat\Psi$ from it (\cref{fig:algorithm}\letterimagetext{e});
    \item apply the wave-packet creation operator $\hat\Psi$ to the vacuum (\cref{fig:algorithm}\letterimagetext{g}) of a large system of size $L \gg l$ (length scale of green-contour panels in \cref{fig:algorithm}), found e.g., using density matrix renormalization group (DMRG) \cite{White1992DensityMatrixFormulation,
    White1993DensityMatrixAlgorithmsQuantum,
    Schollwock2005DensitymatrixRenormalizationGroup} 
    (\cref{fig:algorithm}\letterimagetext{f}), to produce the input scattering state.
\end{enumerate}
These steps define a hierarchy of length scales $\ell_\mathcal{W} \ll l \ll L$, which is valid for sufficiently large $l$ and $L$ if the system is in a gapped
phase, and the single-quasiparticle band is spectrally distinguishable (also by decoupling symmetry sectors).
This hierarchy of length scales naturally holds in homogeneous gapped phases, where low-energy quasiparticle excitations are infrared objects controlled by a finite correlation length $\xi$. We conjecture that for $l \gg \xi$ the intermediate system already captures the bulk structure, while $L \gg l$ suppresses boundary effects \cite{hastingsSpectralGapExponential2006}.

If the quasiparticle excitations have non-trivial topological content, the algorithm should be modified accordingly; this extension is left for future work.

\subsection{Exact diagonalization at intermediate size}

\label{subsec:spectral}

The goal of this step is to reconstruct the quasiparticle spectrum at intermediate size $l$. To this end, we simultaneously diagonalize the Hamiltonian $\hat{H}$, the translation operator $\hat{T}$, and, when available, a spatial reflection operator $\hat{P}$, thereby assigning energy, (quasi)momentum, and parity quantum numbers to each state.

We represent a quantum state of the chain (see details in Appendix \ref{app:tnmethods}) as an $l$-leg tensor, with each leg having dimension $d$:
\begin{equation}
|\psi \rangle =\tikzset{every picture/.style={line width=0.75pt}} 
\begin{tikzpicture}[x=0.75pt,y=0.75pt,yscale=-1,xscale=1, baseline=(XXXX.south) ]
\path (0,51);\path (91.28385416666669,0);\draw    ($(current bounding box.center)+(0,0.3em)$) node [anchor=south] (XXXX) {};
\draw [color={rgb, 255:red, 0; green, 0; blue, 0 }  ,draw opacity=1 ]   (20,25) -- (20,35) ;
\draw [color={rgb, 255:red, 0; green, 0; blue, 0 }  ,draw opacity=1 ]   (9.83,25) -- (9.83,35) ;
\draw  [color={rgb, 255:red, 0; green, 0; blue, 0 }  ,draw opacity=1 ][fill={rgb, 255:red, 255; green, 255; blue, 255 }  ,fill opacity=1 ] (5,16.83) .. controls (5,13.06) and (8.06,10) .. (11.83,10) -- (79.17,10) .. controls (82.94,10) and (86,13.06) .. (86,16.83) -- (86,18.17) .. controls (86,21.94) and (82.94,25) .. (79.17,25) -- (11.83,25) .. controls (8.06,25) and (5,21.94) .. (5,18.17) -- cycle ;
\draw [color={rgb, 255:red, 0; green, 0; blue, 0 }  ,draw opacity=1 ]   (30,25) -- (30,35) ;
\draw [color={rgb, 255:red, 0; green, 0; blue, 0 }  ,draw opacity=1 ]   (80,25) -- (80,35) ;
\draw [color={rgb, 255:red, 0; green, 0; blue, 0 }  ,draw opacity=1 ]   (70,25) -- (70,35) ;
\draw [color={rgb, 255:red, 0; green, 0; blue, 0 }  ,draw opacity=1 ]   (60,25) -- (60,35) ;
\draw (9.38,41.3) node  [font=\scriptsize]  {$1$};
\draw (45.5,17.5) node  [font=\footnotesize]  {$\psi $};
\draw (19.38,41.3) node  [font=\scriptsize]  {$2$};
\draw (79.38,42.3) node  [font=\scriptsize]  {$l$};
\draw (31.09,43.93) node [anchor=south] [inner sep=0.75pt]  [font=\scriptsize]  {$...$};
\draw (46.09,33.32) node [anchor=south] [inner sep=0.75pt]  [font=\scriptsize]  {$...$};
\end{tikzpicture}
\ .
\end{equation}
By construction, there exists a unitary translation operator $\hat{T}$ which is an $l$-root of unity,
\begin{math}
\hat{T}^{l} =\mathbb{I}
\end{math},
and whose eigenvalues are thus complex phases $e^{2\pi n/l} ,\ n\in \{0,\dotsc ,l-1\}$.

We can also construct a spatial reflection (parity) operator $\hat{P}$, which is unitary and involutory (and so Hermitian) and mirrors the action of the translation:
\begin{equation}
\hat{P}^{2} =\mathbb{I} \ ,\quad \hat{P}\hat{T}\hat{P} 
=\hat{T}^{\dagger } \ .
\end{equation}
We assume the system Hamiltonian $\hat{H}$ to be geometrically $\ell_{h}$-local, i.e., it is decomposable as
\begin{equation}
\hat{H} =\sum _{j=1}^{l}\hat{h}_{j} =\sum _{j=1}^{l}
\tikzset{every picture/.style={line width=0.75pt}} 
\begin{tikzpicture}[x=0.75pt,y=0.75pt,yscale=-1,xscale=1, baseline=(XXXX.south) ]
\path (0,66);\path (111.15885416666663,0);\draw    ($(current bounding box.center)+(0,0.3em)$) node [anchor=south] (XXXX) {};
\draw [color={rgb, 255:red, 0; green, 0; blue, 0 }  ,draw opacity=1 ]   (45,15) -- (45,50) ;
\draw [color={rgb, 255:red, 0; green, 0; blue, 0 }  ,draw opacity=1 ]   (55,15) -- (55,50) ;
\draw [color={rgb, 255:red, 0; green, 0; blue, 0 }  ,draw opacity=1 ]   (15,15) -- (15,50) ;
\draw [color={rgb, 255:red, 0; green, 0; blue, 0 }  ,draw opacity=1 ]   (5,15) -- (5,50) ;
\draw [color={rgb, 255:red, 0; green, 0; blue, 0 }  ,draw opacity=1 ]   (25,15) -- (25,50) ;
\draw [color={rgb, 255:red, 0; green, 0; blue, 0 }  ,draw opacity=1 ]   (35,15) -- (35,50) ;
\draw  [color={rgb, 255:red, 0; green, 0; blue, 0 }  ,draw opacity=1 ][fill={rgb, 255:red, 255; green, 255; blue, 255 }  ,fill opacity=1 ] (30,26.49) .. controls (30,25.67) and (30.67,25) .. (31.49,25) -- (58.51,25) .. controls (59.33,25) and (60,25.67) .. (60,26.49) -- (60,38.51) .. controls (60,39.33) and (59.33,40) .. (58.51,40) -- (31.49,40) .. controls (30.67,40) and (30,39.33) .. (30,38.51) -- cycle ;
\draw [color={rgb, 255:red, 0; green, 0; blue, 0 }  ,draw opacity=1 ]   (75,15) -- (75,50) ;
\draw [color={rgb, 255:red, 0; green, 0; blue, 0 }  ,draw opacity=1 ]   (65,15) -- (65,50) ;
\draw [color={rgb, 255:red, 0; green, 0; blue, 0 }  ,draw opacity=1 ]   (85,15) -- (85,50) ;
\draw [color={rgb, 255:red, 0; green, 0; blue, 0 }  ,draw opacity=1 ]   (95,15) -- (95,50) ;
\draw [color={rgb, 255:red, 0; green, 0; blue, 0 }  ,draw opacity=1 ]   (105,15) -- (105,50) ;
\draw (44.38,54.96) node  [font=\scriptsize]  {$j$};
\draw (45,32.5) node  [font=\footnotesize]  {$h$};
\draw (4.71,56.3) node  [font=\scriptsize]  {$1$};
\draw (104.38,56.3) node  [font=\scriptsize]  {$l$};
\end{tikzpicture}
,
\end{equation}
where all the local Hamiltonians are obtained by translations $\hat{T}^{\dagger }\hat{h}_{j}\hat{T} =\hat{h}_{j+1}$. 
Often, we can consider systems in which parity symmetry holds for the local Hamiltonian, i.e., $\hat{P}\hat{h}_{j}\hat{P} =\hat{h}_{l-j}$.
From translation and reflection invariance of $\hat{h}$, it follows straightforwardly that $\hat{H}$ has translation and parity symmetry:
\begin{equation}
[\hat{H} ,\hat{T}] =0\ ,\quad [\hat{H} ,\hat{P}] =0\ .
\end{equation}
The pair $\hat{H}, \hat{T}$ forms a set of commuting observables, so it admits a basis of common eigenvectors, the Bloch states $|\omega ,k\rangle $:
\begin{equation}
\hat{H} |\omega ,k\rangle = \omega |\omega 
,k\rangle \ ,\quad \hat{T} |\omega ,k\rangle =e^{ik} 
|\omega ,k\rangle \ ,
\end{equation}
where we omit the internal degeneracy indices, if any.

Since $\hat{T}$ and $\hat{P}$ together generate the dihedral group $D_l$ (which is non-Abelian), they cannot be simultaneously diagonalized at a generic (quasi)momentum $k$. 
A simultaneous eigenbasis of $\hat{H}$, $\hat{T}$, and $\hat{P}$ exists only at the inversion-invariant momenta $k=0$ and $k=\pi$, where the parity quantum number is well defined.
Indeed, parity acts on the Bloch states as follows
\begin{equation}
\hat{P} |\omega ,k\rangle =\alpha _{k} |\omega
,-k\rangle \ ,
\end{equation}
where $\alpha _{k}$ is a phase which allows symmetrization or anti-symmetrization of Bloch states (see details in Appendix \ref{app:wannierfunctions}).

In addition to $\hat{T}$ and $\hat{P}$, the system may possess global pointwise symmetries, i.e.\ operators of the form $\hat{C} = \hat{c} \otimes \hat{c} \otimes \cdots \otimes \hat{c}$, where $\hat{c}$ acts identically on each site. 
By construction, such symmetries commute with $\hat{T}$ and $\hat{P}$, so the Hilbert space further decomposes into symmetry sectors.

For simplicity, we consider a system with a unique ground-state vacuum $|\Omega\rangle$. The presence of a false vacuum or even spontaneously symmetry-breaking degenerate ground states would not qualitatively affect the algorithm and could be accommodated with only minor modifications.

In the spectrum $(\omega, k)$ we are looking for an excitation whose band can be spectrally decoupled, possibly removing degeneracies and overlaps by restricting the whole spectrum to a specific symmetry sector; if, after this restriction, this band takes the form of a single isolated quasiparticle dispersion relation $\omega_k$, it defines a stable single-quasiparticle $\phi$, where $\phi$ is the quasiparticle label (i.e., the quasiparticle species).
In this regime, Wannier localization can be performed as described in \cref{subsec:wannierloc}.
If two quasiparticle bands are intersecting even after symmetry sector restrictions (same quantum numbers), the Wannier quasiparticle localization is still possible but requires generalization as in \cite{Marzari1997MaximallyLocalizedGeneralized, Marzari2012MaximallyLocalizedWannier}, which is material for future work.

For intermediate system sizes $l=\mathcal{O}\left( 10^{1}\right)$, the Bloch states of the band $|\phi _{k} \rangle$ are assumed to be computable using an available technique.
The one used in this paper is mainly (a) ED methods, $\dim\mathcal{H} =d^{l} =\mathcal{O}\left( 10^{4}\right)$ and (b) Krylov methods, which easily reach $\dim\mathcal{H} = \mathcal{O}\left( 10^{6}\right)$, and allows us to consider systems of length $l=\mathcal{O}\left( 10^{1}\right)$ for small local dimensions such as $d \leq 3$, which is the case of interest in this paper.
When $\hat{H}$ is parity symmetric, we can optimize the ED step by restricting the spectrum to non-negative momentum states.

To each quasiparticle $\phi$ are assigned all the quantum numbers associated with each symmetry, i.e., space-time and (pointwise) internal symmetries.
If the symmetry group is non-Abelian, the label will correspond to an irreducible representation (irrep), with a primary quantum number, labeling the quasiparticle species, and a secondary one, labeling internal quasiparticle states.

\subsection{Wannier state localization}

\label{subsec:wannierloc}

In order to build compact real-space wave-packet states, we need to localize the single-quasiparticle excitations. 
We can address this problem using the basis of Wannier states:
\begin{equation}
|\phi _{\theta } \rangle =\frac{1}{\sqrt{l}}\sum _{k} e^{i\theta _{k}} |\phi _{k} \rangle \ ,
\end{equation}
where $e^{i\theta _{k}}$ are $l$ arbitrary phases, one for each single-quasiparticle Bloch momentum state \cite{wannierStructureElectronicExcitation1937,Wannier1962DynamicsBandElectrons}.
One can prove (see Appendix \ref{app:wannierfunctions}) that all the translated states $\hat{T}^{j} |\phi _{\theta } \rangle $ form an orthonormal basis, like the Bloch states.

We define the local excess energy density above the ground state\footnote{Notice that, since there is no unique choice of $\hat h_j$ which produces the same $\hat H$, the energy excess \eqref{eq:enexcess} is not 
uniquely defined. However, any choices differ only by a local redistribution that do not affect the localization process 
qualitatively; thus, in general, we choose $\hat h_j$ with 
minimum support, which is compatible with all the 
symmetries of the problem.}
\begin{equation}
\varepsilon _{j} =\langle \phi _{\theta } |\hat{h}_{j} 
|\phi _{\theta } \rangle -\langle \Omega_l |\hat{h}_{j} 
|\Omega_l \rangle \ , \label{eq:enexcess}
\end{equation}
and we construct a positive definite, normalized distribution from it
\begin{equation}
p_{j}[ \theta ] =\frac{| \varepsilon _{j}| }{\sum _{j}| 
\varepsilon _{j}| } \ ,\ \label{eq:wannierprobability}
\end{equation}
which sums to one by construction and can be interpreted as an energy-based probability density associated with the single quasiparticle.
Given a distance function $x_{j}$ from a given site $j_{0}$ (e.g., the central site), such as $x_{j}^{2} \sim \sin^{2}( \pi ( j-j_{0}) /l)$, we can construct a spread functional corresponding to an energy-based localization measure from the second moment of $p_{j}$:
\begin{equation}
\sigma ^{2}[ \theta ] =\sum _{j=1}^{l} x_{j}^{2} p_{j}
[ \theta ] \ .\ \label{eq:sigma2functional}
\end{equation}
This functional plays the role of the Marzari--Vanderbilt functional in the Wannier function minimization schemes of \cite{Marzari1997MaximallyLocalizedGeneralized,Marzari2012MaximallyLocalizedWannier}. 
The functional of \cref{eq:sigma2functional} can be numerically minimized in the compact domain $\{\theta _{k}\}\in[ -\pi ,\pi ]^{l}$.
The optimum $\theta ^{\star }$ provides the so-called maximally localized Wannier function (MLWF) around the site $j_0$, 
\begin{equation}
|\phi _{j_{0}} \rangle \equiv 
|\phi _{\theta ^{\star }} \rangle 
\ , \quad
\sigma ^{2}\left[ \theta ^{\star }\right] 
=\underset{\theta }{\min} \thinspace \sigma ^{2}[ \theta ] 
\ .
\end{equation}
Notice that this energy localization of the Wannier states is not the standard geometric localization in \cite{Marzari1997MaximallyLocalizedGeneralized,Marzari2012MaximallyLocalizedWannier}. 
The minimization can be done easily after simplifying the expression \eqref{eq:sigma2functional}, see Appendix \ref{app:wannierfunctions} for details. 

It was proven by Kohn \cite{Kohn1959AnalyticPropertiesBloch} that for a one-dimensional periodic, isolated, non-degenerate Bloch band (a tight-binding regime is a sufficient condition \cite{ashcroftSolidStatePhysics1976}) there exists exactly one (up to an overall sign) real Wannier function that is symmetric (or antisymmetric) under reflection and decays exponentially with distance. 
This implies that the density $p_{j}$ associated to the MLWF $|\phi _{j_{0}} \rangle $ decays exponentially with the distance $x_{j}$ from $j_{0}$ (details in Appendix \ref{app:wannierfunctions}). 
For computational purposes, given a numerical threshold $\epsilon \ll 1$, we refer to
\begin{equation}\label{eq:wannier_support}
\mathcal{W} =\left\{j:\ p_{j}
\left[ \theta ^{\star }\right]  >\epsilon \right\}
\end{equation}
as the support of the Wannier localized excitation, of length $\ell_\mathcal{W} = |\mathcal{W}| = 2r+1$, 
where $r$ is the Wannier radius. 
In graphical notation, the Wannier state can be represented as an $l$-site vector with $\ell_\mathcal{W} $ sites belonging to the Wannier support (see also \cref{fig:algorithm}):
\begin{equation}
|\phi_{j} \rangle =\tikzset{every picture/.style={line width=0.75pt}} 
\begin{tikzpicture}[x=0.75pt,y=0.75pt,yscale=-1,xscale=1, baseline=(XXXX.south) ]
\path (0,73);\path (136.609375,0);\draw    ($(current bounding box.center)+(0,0.3em)$) node [anchor=south] (XXXX) {};
\draw [color={rgb, 255:red, 0; green, 0; blue, 0 }  ,draw opacity=1 ]   (20,36) -- (20,46) ;
\draw [color={rgb, 255:red, 0; green, 0; blue, 0 }  ,draw opacity=1 ]   (9.83,36) -- (9.83,46) ;
\draw  [color={rgb, 255:red, 0; green, 0; blue, 0 }  ,draw opacity=1 ][fill={rgb, 255:red, 255; green, 255; blue, 255 }  ,fill opacity=1 ] (5,27.83) .. controls (5,24.06) and (8.06,21) .. (11.83,21) -- (123.17,21) .. controls (126.94,21) and (130,24.06) .. (130,27.83) -- (130,29.17) .. controls (130,32.94) and (126.94,36) .. (123.17,36) -- (11.83,36) .. controls (8.06,36) and (5,32.94) .. (5,29.17) -- cycle ;
\draw [color={rgb, 255:red, 0; green, 0; blue, 0 }  ,draw opacity=1 ]   (115,36) -- (115,46) ;
\draw [color={rgb, 255:red, 0; green, 0; blue, 0 }  ,draw opacity=1 ]   (70,36) -- (70,46) ;
\draw [color={rgb, 255:red, 0; green, 0; blue, 0 }  ,draw opacity=1 ]   (80,36) -- (80,46) ;
\draw [color={rgb, 255:red, 0; green, 0; blue, 0 }  ,draw opacity=1 ]   (60,36) -- (60,46) ;
\draw [color={rgb, 255:red, 0; green, 0; blue, 0 }  ,draw opacity=1 ]   (125,36) -- (125,46) ;
\draw [color={rgb, 255:red, 0; green, 0; blue, 0 }  ,draw opacity=1 ]   (50,36) -- (50,46) ;
\draw [color={rgb, 255:red, 0; green, 0; blue, 0 }  ,draw opacity=1 ]   (90,36) -- (90,46) ;
\draw    (50,51) -- (90,51) ;
\draw [shift={(90,51)}, rotate = 180] [color={rgb, 255:red, 0; green, 0; blue, 0 }  ][line width=0.75]    (0,3.35) -- (0,-3.35)(6.56,-1.97) .. controls (4.17,-0.84) and (1.99,-0.18) .. (0,0) .. controls (1.99,0.18) and (4.17,0.84) .. (6.56,1.97)   ;
\draw [shift={(50,51)}, rotate = 0] [color={rgb, 255:red, 0; green, 0; blue, 0 }  ][line width=0.75]    (0,3.35) -- (0,-3.35)(6.56,-1.97) .. controls (4.17,-0.84) and (1.99,-0.18) .. (0,0) .. controls (1.99,0.18) and (4.17,0.84) .. (6.56,1.97)   ;
\draw (10,52) node  [font=\scriptsize]  {$1$};
\draw (67.5,28.5) node  [font=\scriptsize]  {$\phi _{j}$};
\draw (20,52) node  [font=\scriptsize]  {$2$};
\draw (124,52) node  [font=\scriptsize]  {$l$};
\draw (28,39.5) node [anchor=north west][inner sep=0.75pt]  [font=\scriptsize]  {$\dotsc $};
\draw (68,61) node  [font=\scriptsize]  {$\ell_\mathcal{W}$};
\draw (98,39.5) node [anchor=north west][inner sep=0.75pt]  [font=\scriptsize]  {$\dotsc $};
\end{tikzpicture}
\ .
\end{equation}
\subsection{The unitary dressed creation operator}

\label{subsec:dressedmpo}

We now describe a procedure to obtain a creation operator for the quasiparticle $\phi$.
We look for an operator $\hat{\phi }_{j}^{\dagger }$ with support $\mathcal{W}$ as in \cref{eq:wannier_support}, that intertwines between the vacuum and the MLWF with maximum fidelity:
\begin{equation}
\label{eq:procrustes}
F = \max_{\hat{\phi}_{j}^{\dagger}} |\langle\phi_{j}| \hat{\phi}_{j}^{\dagger} |\Omega_l \rangle|^2
\quad\text{s.t.}\quad
\lVert \hat{\phi}_{j}^{\dagger} |\Omega_l \rangle \rVert = 1
\ .
\end{equation}
We enforce the norm constraint by imposing the stronger condition that $\hat{\phi }^{\dagger }$ is unitary, so that it may implemented directly as a quantum circuit by gate decomposition.
The optimization problem then becomes a variant of the orthogonal Procrustes problem \cite{schonemannGeneralizedSolutionOrthogonal1966}, which can be solved as follows (see Appendix \ref{app:wavepacket} for the proof):
\begin{enumerate}
    \item 
compute the following partial trace outside the Wannier support
\begin{equation}
\hat{A}^\dagger =\Tr_{\overline{\mathcal{W}}} |\phi _{j} 
\rangle \langle \Omega_l |=\tikzset{every picture/.style={line width=0.75pt}} 
\begin{tikzpicture}[x=0.75pt,y=0.75pt,yscale=-1,xscale=1, baseline=(XXXX.south) ]
\path (0,84);\path (142.609375,0);\draw    ($(current bounding box.center)+(0,0.3em)$) node [anchor=south] (XXXX) {};
\draw  [color={rgb, 255:red, 0; green, 0; blue, 0 }  ,draw opacity=1 ][fill={rgb, 255:red, 255; green, 255; blue, 255 }  ,fill opacity=1 ] (5,51.83) .. controls (5,48.06) and (8.06,45) .. (11.83,45) -- (128.17,45) .. controls (131.94,45) and (135,48.06) .. (135,51.83) -- (135,53.17) .. controls (135,56.94) and (131.94,60) .. (128.17,60) -- (11.83,60) .. controls (8.06,60) and (5,56.94) .. (5,53.17) -- cycle ;
\draw [color={rgb, 255:red, 0; green, 0; blue, 0 }  ,draw opacity=1 ]   (70,60) -- (70,70) ;
\draw [color={rgb, 255:red, 0; green, 0; blue, 0 }  ,draw opacity=1 ]   (80,60) -- (80,70) ;
\draw [color={rgb, 255:red, 0; green, 0; blue, 0 }  ,draw opacity=1 ]   (60,60) -- (60,70) ;
\draw [color={rgb, 255:red, 0; green, 0; blue, 0 }  ,draw opacity=1 ]   (50,60) -- (50,70) ;
\draw [color={rgb, 255:red, 0; green, 0; blue, 0 }  ,draw opacity=1 ]   (90,60) -- (90,70) ;
\draw  [color={rgb, 255:red, 0; green, 0; blue, 0 }  ,draw opacity=1 ][fill={rgb, 255:red, 255; green, 255; blue, 255 }  ,fill opacity=1 ] (5,31.83) .. controls (5,28.06) and (8.06,25) .. (11.83,25) -- (128.17,25) .. controls (131.94,25) and (135,28.06) .. (135,31.83) -- (135,33.17) .. controls (135,36.94) and (131.94,40) .. (128.17,40) -- (11.83,40) .. controls (8.06,40) and (5,36.94) .. (5,33.17) -- cycle ;
\draw    (20,25) .. controls (8.66,3.37) and (8.99,79.87) .. (20,60) ;
\draw    (10,25) .. controls (-1.34,3.37) and (-1.01,79.87) .. (10,60) ;
\draw    (40,25) .. controls (28.66,3.37) and (28.99,79.87) .. (40,60) ;
\draw    (130,25) .. controls (141.34,3.37) and (141.01,79.87) .. (130,60) ;
\draw    (120,25) .. controls (131.34,3.37) and (131.01,79.87) .. (120,60) ;
\draw    (100,25) .. controls (111.34,3.37) and (111.01,79.87) .. (100,60) ;
\draw [color={rgb, 255:red, 0; green, 0; blue, 0 }  ,draw opacity=1 ]   (70,15) -- (70,25) ;
\draw [color={rgb, 255:red, 0; green, 0; blue, 0 }  ,draw opacity=1 ]   (80,15) -- (80,25) ;
\draw [color={rgb, 255:red, 0; green, 0; blue, 0 }  ,draw opacity=1 ]   (60,15) -- (60,25) ;
\draw [color={rgb, 255:red, 0; green, 0; blue, 0 }  ,draw opacity=1 ]   (50,15) -- (50,25) ;
\draw [color={rgb, 255:red, 0; green, 0; blue, 0 }  ,draw opacity=1 ]   (90,15) -- (90,25) ;
\draw (70,32.5) node  [font=\footnotesize]  {$\Omega_l $};
\draw (69,76) node  [font=\scriptsize]  {$\mathcal{W}$};
\draw (70,52.5) node  [font=\footnotesize]  {$\phi_{j}$};
\end{tikzpicture}
\ ;
\end{equation}
\item
perform a singular value decomposition (SVD)
\begin{equation}
\hat{A}^\dag =\tikzset{every picture/.style={line width=0.75pt}} 
\begin{tikzpicture}[x=0.75pt,y=0.75pt,yscale=-1,xscale=1, baseline=(XXXX.south) ]
\path (0,46);\path (79.92447916666669,0);\draw    ($(current bounding box.center)+(0,0.3em)$) node [anchor=south] (XXXX) {};
\draw  [color={rgb, 255:red, 0; green, 0; blue, 0 }  ,draw opacity=1 ][fill={rgb, 255:red, 255; green, 255; blue, 255 }  ,fill opacity=1 ] (15,17.36) .. controls (15,16.06) and (16.06,15) .. (17.36,15) -- (62.64,15) .. controls (63.94,15) and (65,16.06) .. (65,17.36) -- (65,27.64) .. controls (65,28.94) and (63.94,30) .. (62.64,30) -- (17.36,30) .. controls (16.06,30) and (15,28.94) .. (15,27.64) -- cycle ;
\draw [color={rgb, 255:red, 0; green, 0; blue, 0 }  ,draw opacity=1 ]   (40,5) -- (40,15) ;
\draw [color={rgb, 255:red, 0; green, 0; blue, 0 }  ,draw opacity=1 ]   (50,5) -- (50,15) ;
\draw [color={rgb, 255:red, 0; green, 0; blue, 0 }  ,draw opacity=1 ]   (30,5) -- (30,15) ;
\draw [color={rgb, 255:red, 0; green, 0; blue, 0 }  ,draw opacity=1 ]   (20.5,5) -- (20.5,15) ;
\draw [color={rgb, 255:red, 0; green, 0; blue, 0 }  ,draw opacity=1 ]   (60,5) -- (60,15) ;
\draw [color={rgb, 255:red, 0; green, 0; blue, 0 }  ,draw opacity=1 ]   (40,30) -- (40,40) ;
\draw [color={rgb, 255:red, 0; green, 0; blue, 0 }  ,draw opacity=1 ]   (50,30) -- (50,40) ;
\draw [color={rgb, 255:red, 0; green, 0; blue, 0 }  ,draw opacity=1 ]   (30,30) -- (30,40) ;
\draw [color={rgb, 255:red, 0; green, 0; blue, 0 }  ,draw opacity=1 ]   (20.5,30) -- (20.5,40) ;
\draw [color={rgb, 255:red, 0; green, 0; blue, 0 }  ,draw opacity=1 ]   (60,30) -- (60,40) ;
\draw [color={rgb, 255:red, 167; green, 167; blue, 167 }  ,draw opacity=1 ] [dash pattern={on 1.5pt off 1.5pt}]  (5,23) -- (75,23) ;
\draw (40,22.5) node  [font=\footnotesize]  {$A^\dagger$};
\end{tikzpicture}
=\tikzset{every picture/.style={line width=0.75pt}} 
\begin{tikzpicture}[x=0.75pt,y=0.75pt,yscale=-1,xscale=1, baseline=(XXXX.south) ]
\path (0,81);\path (60.265625,0);\draw    ($(current bounding box.center)+(0,0.3em)$) node [anchor=south] (XXXX) {};
\draw  [color={rgb, 255:red, 0; green, 0; blue, 0 }  ,draw opacity=1 ][fill={rgb, 255:red, 255; green, 255; blue, 255 }  ,fill opacity=1 ] (5,21.83) .. controls (5,18.06) and (8.06,15) .. (11.83,15) -- (48.17,15) .. controls (51.94,15) and (55,18.06) .. (55,21.83) -- (55,23.17) .. controls (55,26.94) and (51.94,30) .. (48.17,30) -- (11.83,30) .. controls (8.06,30) and (5,26.94) .. (5,23.17) -- cycle ;
\draw [color={rgb, 255:red, 0; green, 0; blue, 0 }  ,draw opacity=1 ]   (30,5) -- (30,15) ;
\draw [color={rgb, 255:red, 0; green, 0; blue, 0 }  ,draw opacity=1 ]   (40,5) -- (40,15) ;
\draw [color={rgb, 255:red, 0; green, 0; blue, 0 }  ,draw opacity=1 ]   (20,5) -- (20,15) ;
\draw [color={rgb, 255:red, 0; green, 0; blue, 0 }  ,draw opacity=1 ]   (10.5,5) -- (10.5,15) ;
\draw [color={rgb, 255:red, 0; green, 0; blue, 0 }  ,draw opacity=1 ]   (50,5) -- (50,15) ;
\draw [color={rgb, 255:red, 0; green, 0; blue, 0 }  ,draw opacity=1 ][line width=1.5]    (30,30) -- (30,50) ;
\draw  [color={rgb, 255:red, 0; green, 0; blue, 0 }  ,draw opacity=1 ][fill={rgb, 255:red, 255; green, 255; blue, 255 }  ,fill opacity=1 ] (5,56.83) .. controls (5,53.06) and (8.06,50) .. (11.83,50) -- (48.17,50) .. controls (51.94,50) and (55,53.06) .. (55,56.83) -- (55,58.17) .. controls (55,61.94) and (51.94,65) .. (48.17,65) -- (11.83,65) .. controls (8.06,65) and (5,61.94) .. (5,58.17) -- cycle ;
\draw [color={rgb, 255:red, 0; green, 0; blue, 0 }  ,draw opacity=1 ]   (30,65) -- (30,75) ;
\draw [color={rgb, 255:red, 0; green, 0; blue, 0 }  ,draw opacity=1 ]   (40,65) -- (40,75) ;
\draw [color={rgb, 255:red, 0; green, 0; blue, 0 }  ,draw opacity=1 ]   (20,65) -- (20,75) ;
\draw [color={rgb, 255:red, 0; green, 0; blue, 0 }  ,draw opacity=1 ]   (10.5,65) -- (10.5,75) ;
\draw [color={rgb, 255:red, 0; green, 0; blue, 0 }  ,draw opacity=1 ]   (50,65) -- (50,75) ;
\draw [color={rgb, 255:red, 0; green, 0; blue, 0 }  ,draw opacity=1 ]   (30,40) ;
\draw  [fill={rgb, 255:red, 255; green, 255; blue, 255 }  ,fill opacity=1 ] (26.75,39.88) .. controls (26.75,38.15) and (28.15,36.75) .. (29.88,36.75) .. controls (31.6,36.75) and (33,38.15) .. (33,39.88) .. controls (33,41.6) and (31.6,43) .. (29.88,43) .. controls (28.15,43) and (26.75,41.6) .. (26.75,39.88) -- cycle ;
\draw (30,22.5) node  [font=\footnotesize]  {$V$};
\draw (38.33,40.17) node  [font=\footnotesize]  {$\Sigma $};
\draw (30,57.5) node  [font=\footnotesize]  {$U$};
\end{tikzpicture}
=\sum _{n} \sigma _{n} |u_{n} \rangle \langle v_{n} |\ ;
\end{equation}
\item
rescale all the singular values to one
\begin{equation}
\hat{\phi }^{\dagger } =\sum _{n} |u_{n} \rangle \langle 
    v_{n} |=\tikzset{every picture/.style={line width=0.75pt}} 
\begin{tikzpicture}[x=0.75pt,y=0.75pt,yscale=-1,xscale=1, baseline=(XXXX.south) ]
\path (0,71);\path (60.265625,0);\draw    ($(current bounding box.center)+(0,0.3em)$) node [anchor=south] (XXXX) {};
\draw  [color={rgb, 255:red, 0; green, 0; blue, 0 }  ,draw opacity=1 ][fill={rgb, 255:red, 255; green, 255; blue, 255 }  ,fill opacity=1 ] (5,21.83) .. controls (5,18.06) and (8.06,15) .. (11.83,15) -- (48.17,15) .. controls (51.94,15) and (55,18.06) .. (55,21.83) -- (55,23.17) .. controls (55,26.94) and (51.94,30) .. (48.17,30) -- (11.83,30) .. controls (8.06,30) and (5,26.94) .. (5,23.17) -- cycle ;
\draw [color={rgb, 255:red, 0; green, 0; blue, 0 }  ,draw opacity=1 ]   (30,5) -- (30,15) ;
\draw [color={rgb, 255:red, 0; green, 0; blue, 0 }  ,draw opacity=1 ]   (40,5) -- (40,15) ;
\draw [color={rgb, 255:red, 0; green, 0; blue, 0 }  ,draw opacity=1 ]   (20,5) -- (20,15) ;
\draw [color={rgb, 255:red, 0; green, 0; blue, 0 }  ,draw opacity=1 ]   (10.5,5) -- (10.5,15) ;
\draw [color={rgb, 255:red, 0; green, 0; blue, 0 }  ,draw opacity=1 ]   (50,5) -- (50,15) ;
\draw [color={rgb, 255:red, 0; green, 0; blue, 0 }  ,draw opacity=1 ][line width=1.5]    (30,30) -- (30,40) ;
\draw  [color={rgb, 255:red, 0; green, 0; blue, 0 }  ,draw opacity=1 ][fill={rgb, 255:red, 255; green, 255; blue, 255 }  ,fill opacity=1 ] (5,46.83) .. controls (5,43.06) and (8.06,40) .. (11.83,40) -- (48.17,40) .. controls (51.94,40) and (55,43.06) .. (55,46.83) -- (55,48.17) .. controls (55,51.94) and (51.94,55) .. (48.17,55) -- (11.83,55) .. controls (8.06,55) and (5,51.94) .. (5,48.17) -- cycle ;
\draw [color={rgb, 255:red, 0; green, 0; blue, 0 }  ,draw opacity=1 ]   (30,55) -- (30,65) ;
\draw [color={rgb, 255:red, 0; green, 0; blue, 0 }  ,draw opacity=1 ]   (40,55) -- (40,65) ;
\draw [color={rgb, 255:red, 0; green, 0; blue, 0 }  ,draw opacity=1 ]   (20,55) -- (20,65) ;
\draw [color={rgb, 255:red, 0; green, 0; blue, 0 }  ,draw opacity=1 ]   (10.5,55) -- (10.5,65) ;
\draw [color={rgb, 255:red, 0; green, 0; blue, 0 }  ,draw opacity=1 ]   (50,55) -- (50,65) ;
\draw (30,22.5) node  [font=\footnotesize]  {$V$};
\draw (30,47.5) node  [font=\footnotesize]  {$U$};
\end{tikzpicture}
=\tikzset{every picture/.style={line width=0.75pt}} 
\begin{tikzpicture}[x=0.75pt,y=0.75pt,yscale=-1,xscale=1, baseline=(XXXX.south) ]
\path (0,46);\path (61.588541666666686,0);\draw    ($(current bounding box.center)+(0,0.3em)$) node [anchor=south] (XXXX) {};
\draw  [color={rgb, 255:red, 0; green, 0; blue, 0 }  ,draw opacity=1 ][fill={rgb, 255:red, 255; green, 255; blue, 255 }  ,fill opacity=1 ] (5,17.36) .. controls (5,16.06) and (6.06,15) .. (7.36,15) -- (52.64,15) .. controls (53.94,15) and (55,16.06) .. (55,17.36) -- (55,27.64) .. controls (55,28.94) and (53.94,30) .. (52.64,30) -- (7.36,30) .. controls (6.06,30) and (5,28.94) .. (5,27.64) -- cycle ;
\draw [color={rgb, 255:red, 0; green, 0; blue, 0 }  ,draw opacity=1 ]   (30,5) -- (30,15) ;
\draw [color={rgb, 255:red, 0; green, 0; blue, 0 }  ,draw opacity=1 ]   (40,5) -- (40,15) ;
\draw [color={rgb, 255:red, 0; green, 0; blue, 0 }  ,draw opacity=1 ]   (20,5) -- (20,15) ;
\draw [color={rgb, 255:red, 0; green, 0; blue, 0 }  ,draw opacity=1 ]   (10.5,5) -- (10.5,15) ;
\draw [color={rgb, 255:red, 0; green, 0; blue, 0 }  ,draw opacity=1 ]   (50,5) -- (50,15) ;
\draw [color={rgb, 255:red, 0; green, 0; blue, 0 }  ,draw opacity=1 ]   (30,30) -- (30,40) ;
\draw [color={rgb, 255:red, 0; green, 0; blue, 0 }  ,draw opacity=1 ]   (40,30) -- (40,40) ;
\draw [color={rgb, 255:red, 0; green, 0; blue, 0 }  ,draw opacity=1 ]   (20,30) -- (20,40) ;
\draw [color={rgb, 255:red, 0; green, 0; blue, 0 }  ,draw opacity=1 ]   (10.5,30) -- (10.5,40) ;
\draw [color={rgb, 255:red, 0; green, 0; blue, 0 }  ,draw opacity=1 ]   (50,30) -- (50,40) ;
\draw (30,22.5) node  [font=\footnotesize]  {$\phi ^{\dagger }$};
\end{tikzpicture}
\ .
\end{equation}
\end{enumerate}
The operator $\hat \phi_j^\dagger$ so constructed generates the best approximation of the MLWF from the vacuum:
\begin{equation}
|\phi_{j} \rangle \simeq \hat{\phi}_{j}^{\dagger} |\Omega_l 
\rangle = \tikzset{every picture/.style={line width=0.75pt}} 
\begin{tikzpicture}[x=0.75pt,y=0.75pt,yscale=-1,xscale=1, baseline=(XXXX.south) ]
\path (0,86);\path (136.609375,0);\draw    ($(current bounding box.center)+(0,0.3em)$) node [anchor=south] (XXXX) {};
\draw [color={rgb, 255:red, 0; green, 0; blue, 0 }  ,draw opacity=1 ]   (20,47) -- (20,57) ;
\draw [color={rgb, 255:red, 0; green, 0; blue, 0 }  ,draw opacity=1 ]   (9.83,47) -- (9.83,57) ;
\draw  [color={rgb, 255:red, 0; green, 0; blue, 0 }  ,draw opacity=1 ][fill={rgb, 255:red, 255; green, 255; blue, 255 }  ,fill opacity=1 ] (5,38.83) .. controls (5,35.06) and (8.06,32) .. (11.83,32) -- (123.17,32) .. controls (126.94,32) and (130,35.06) .. (130,38.83) -- (130,40.17) .. controls (130,43.94) and (126.94,47) .. (123.17,47) -- (11.83,47) .. controls (8.06,47) and (5,43.94) .. (5,40.17) -- cycle ;
\draw [color={rgb, 255:red, 0; green, 0; blue, 0 }  ,draw opacity=1 ]   (115,47) -- (115,57) ;
\draw [color={rgb, 255:red, 0; green, 0; blue, 0 }  ,draw opacity=1 ]   (125,47) -- (125,57) ;
\draw  [color={rgb, 255:red, 0; green, 0; blue, 0 }  ,draw opacity=1 ][fill={rgb, 255:red, 255; green, 255; blue, 255 }  ,fill opacity=1 ] (43.33,59.29) .. controls (43.33,57.99) and (44.39,56.93) .. (45.69,56.93) -- (90.97,56.93) .. controls (92.28,56.93) and (93.33,57.99) .. (93.33,59.29) -- (93.33,69.58) .. controls (93.33,70.88) and (92.28,71.93) .. (90.97,71.93) -- (45.69,71.93) .. controls (44.39,71.93) and (43.33,70.88) .. (43.33,69.58) -- cycle ;
\draw [color={rgb, 255:red, 0; green, 0; blue, 0 }  ,draw opacity=1 ]   (68.33,46.93) -- (68.33,56.93) ;
\draw [color={rgb, 255:red, 0; green, 0; blue, 0 }  ,draw opacity=1 ]   (78.33,46.93) -- (78.33,56.93) ;
\draw [color={rgb, 255:red, 0; green, 0; blue, 0 }  ,draw opacity=1 ]   (58.33,46.93) -- (58.33,56.93) ;
\draw [color={rgb, 255:red, 0; green, 0; blue, 0 }  ,draw opacity=1 ]   (48.83,46.93) -- (48.83,56.93) ;
\draw [color={rgb, 255:red, 0; green, 0; blue, 0 }  ,draw opacity=1 ]   (88.33,46.93) -- (88.33,56.93) ;
\draw [color={rgb, 255:red, 0; green, 0; blue, 0 }  ,draw opacity=1 ]   (68.33,71.93) -- (68.33,81.93) ;
\draw [color={rgb, 255:red, 0; green, 0; blue, 0 }  ,draw opacity=1 ]   (78.33,71.93) -- (78.33,81.93) ;
\draw [color={rgb, 255:red, 0; green, 0; blue, 0 }  ,draw opacity=1 ]   (58.33,71.93) -- (58.33,81.93) ;
\draw [color={rgb, 255:red, 0; green, 0; blue, 0 }  ,draw opacity=1 ]   (48.83,71.93) -- (48.83,81.93) ;
\draw [color={rgb, 255:red, 0; green, 0; blue, 0 }  ,draw opacity=1 ]   (88.33,71.93) -- (88.33,81.93) ;
\draw (67.5,39.5) node  [font=\footnotesize]  {$\Omega_l $};
\draw (68.33,64.43) node  [font=\footnotesize]  {$\phi ^{\dagger }$};
\draw (32.82,54.4) node  [font=\scriptsize]  {$\dotsc $};
\draw (102.82,55.07) node  [font=\scriptsize]  {$\dotsc $};
\end{tikzpicture}\ . \quad \label{eq:creationop}
\end{equation}

Let us now introduce a new, larger system of size $L=\mathcal{O}\left( 10^{2}\right)$, this time with open boundary conditions (OBC), so that an MPS approximation of its ground state
\begin{equation}
|\Omega_L \rangle =\tikzset{every picture/.style={line width=0.75pt}} 
\begin{tikzpicture}[x=0.75pt,y=0.75pt,yscale=-1,xscale=1, baseline=(XXXX.south) ]
\path (0,34);\path (135.1015625,0);\draw    ($(current bounding box.center)+(0,0.3em)$) node [anchor=south] (XXXX) {};
\draw  [color={rgb, 255:red, 0; green, 0; blue, 0 }  ,draw opacity=1 ][fill={rgb, 255:red, 255; green, 255; blue, 255 }  ,fill opacity=1 ] (8,14) .. controls (8,9.58) and (11.58,6) .. (16,6) -- (16,6) .. controls (20.42,6) and (24,9.58) .. (24,14) -- (24,14) .. controls (24,18.42) and (20.42,22) .. (16,22) -- (16,22) .. controls (11.58,22) and (8,18.42) .. (8,14) -- cycle ;
\draw [color={rgb, 255:red, 0; green, 0; blue, 0 }  ,draw opacity=1 ]   (16,22) -- (16,30) ;
\draw [color={rgb, 255:red, 0; green, 0; blue, 0 }  ,draw opacity=1 ]   (24,14) -- (32,14) ;
\draw  [color={rgb, 255:red, 0; green, 0; blue, 0 }  ,draw opacity=1 ][fill={rgb, 255:red, 255; green, 255; blue, 255 }  ,fill opacity=1 ] (32,14) .. controls (32,9.58) and (35.58,6) .. (40,6) -- (40,6) .. controls (44.42,6) and (48,9.58) .. (48,14) -- (48,14) .. controls (48,18.42) and (44.42,22) .. (40,22) -- (40,22) .. controls (35.58,22) and (32,18.42) .. (32,14) -- cycle ;
\draw [color={rgb, 255:red, 0; green, 0; blue, 0 }  ,draw opacity=1 ]   (40,22) -- (40,30) ;
\draw [color={rgb, 255:red, 0; green, 0; blue, 0 }  ,draw opacity=1 ]   (48,14) -- (56,14) ;
\draw  [color={rgb, 255:red, 0; green, 0; blue, 0 }  ,draw opacity=1 ][fill={rgb, 255:red, 255; green, 255; blue, 255 }  ,fill opacity=1 ] (56,14) .. controls (56,9.58) and (59.58,6) .. (64,6) -- (64,6) .. controls (68.42,6) and (72,9.58) .. (72,14) -- (72,14) .. controls (72,18.42) and (68.42,22) .. (64,22) -- (64,22) .. controls (59.58,22) and (56,18.42) .. (56,14) -- cycle ;
\draw [color={rgb, 255:red, 0; green, 0; blue, 0 }  ,draw opacity=1 ]   (64,22) -- (64,30) ;
\draw [color={rgb, 255:red, 0; green, 0; blue, 0 }  ,draw opacity=1 ]   (72,14) -- (80,14) ;
\draw [color={rgb, 255:red, 0; green, 0; blue, 0 }  ,draw opacity=1 ]   (106,14) -- (114,14) ;
\draw  [color={rgb, 255:red, 0; green, 0; blue, 0 }  ,draw opacity=1 ][fill={rgb, 255:red, 255; green, 255; blue, 255 }  ,fill opacity=1 ] (114,14) .. controls (114,9.58) and (117.58,6) .. (122,6) -- (122,6) .. controls (126.42,6) and (130,9.58) .. (130,14) -- (130,14) .. controls (130,18.42) and (126.42,22) .. (122,22) -- (122,22) .. controls (117.58,22) and (114,18.42) .. (114,14) -- cycle ;
\draw [color={rgb, 255:red, 0; green, 0; blue, 0 }  ,draw opacity=1 ]   (122,22) -- (122,30) ;
\draw [color={rgb, 255:red, 0; green, 0; blue, 0 }  ,draw opacity=1 ] [dash pattern={on 0.75pt off 0.75pt}]  (80,14) -- (88,14) ;
\draw [color={rgb, 255:red, 0; green, 0; blue, 0 }  ,draw opacity=1 ] [dash pattern={on 0.75pt off 0.75pt}]  (98,14) -- (106,14) ;
\draw (16,14) node  [font=\scriptsize]  {$\Omega _{1}$};
\draw (40,14) node  [font=\scriptsize]  {$\Omega _{2}$};
\draw (64,14) node  [font=\scriptsize]  {$\Omega _{3}$};
\draw (122,14) node  [font=\scriptsize]  {$\Omega _{L}$};
\end{tikzpicture}
\end{equation}
can be efficiently obtained numerically, e.g., using DMRG.
If $L\gg l \gg \xi \sim \ell_\mathcal{W} $, where $\xi$ is the correlation length of the ground state $| \Omega_l \rangle$, both systems are effectively in the thermodynamic limit. Then it is reasonable to expect that boundary effects are negligible in the bulk, so that $|\Phi_{j} \rangle = \hat \phi_{j}^{\dagger} |\Omega_L \rangle$ provides a good approximation for the MLWFs of the large system:
\begin{equation}
|\Phi _{j} \rangle =\tikzset{every picture/.style={line width=0.75pt}} 
\begin{tikzpicture}[x=0.75pt,y=0.75pt,yscale=-1,xscale=1, baseline=(XXXX.south) ]
\path (0,73);\path (242.76822916666669,0);\draw    ($(current bounding box.center)+(0,0.3em)$) node [anchor=south] (XXXX) {};
\draw  [color={rgb, 255:red, 0; green, 0; blue, 0 }  ,draw opacity=1 ][fill={rgb, 255:red, 255; green, 255; blue, 255 }  ,fill opacity=1 ] (5,29) .. controls (5,24.58) and (8.58,21) .. (13,21) -- (13,21) .. controls (17.42,21) and (21,24.58) .. (21,29) -- (21,29) .. controls (21,33.42) and (17.42,37) .. (13,37) -- (13,37) .. controls (8.58,37) and (5,33.42) .. (5,29) -- cycle ;
\draw [color={rgb, 255:red, 0; green, 0; blue, 0 }  ,draw opacity=1 ]   (13,37) -- (13,45) ;
\draw [color={rgb, 255:red, 0; green, 0; blue, 0 }  ,draw opacity=1 ]   (21,29) -- (29,29) ;
\draw  [color={rgb, 255:red, 0; green, 0; blue, 0 }  ,draw opacity=1 ][fill={rgb, 255:red, 255; green, 255; blue, 255 }  ,fill opacity=1 ] (29,29) .. controls (29,24.58) and (32.58,21) .. (37,21) -- (37,21) .. controls (41.42,21) and (45,24.58) .. (45,29) -- (45,29) .. controls (45,33.42) and (41.42,37) .. (37,37) -- (37,37) .. controls (32.58,37) and (29,33.42) .. (29,29) -- cycle ;
\draw [color={rgb, 255:red, 0; green, 0; blue, 0 }  ,draw opacity=1 ]   (37,37) -- (37,45) ;
\draw [color={rgb, 255:red, 0; green, 0; blue, 0 }  ,draw opacity=1 ]   (45,29) -- (53,29) ;
\draw [color={rgb, 255:red, 0; green, 0; blue, 0 }  ,draw opacity=1 ]   (69,29) -- (77,29) ;
\draw  [color={rgb, 255:red, 0; green, 0; blue, 0 }  ,draw opacity=1 ][fill={rgb, 255:red, 255; green, 255; blue, 255 }  ,fill opacity=1 ] (77,29) .. controls (77,24.58) and (80.58,21) .. (85,21) -- (85,21) .. controls (89.42,21) and (93,24.58) .. (93,29) -- (93,29) .. controls (93,33.42) and (89.42,37) .. (85,37) -- (85,37) .. controls (80.58,37) and (77,33.42) .. (77,29) -- cycle ;
\draw [color={rgb, 255:red, 0; green, 0; blue, 0 }  ,draw opacity=1 ]   (85,37) -- (85,45) ;
\draw [color={rgb, 255:red, 0; green, 0; blue, 0 }  ,draw opacity=1 ]   (93,29) -- (101,29) ;
\draw  [color={rgb, 255:red, 0; green, 0; blue, 0 }  ,draw opacity=1 ][fill={rgb, 255:red, 255; green, 255; blue, 255 }  ,fill opacity=1 ] (77,47.52) .. controls (77,46.13) and (78.13,45) .. (79.52,45) -- (186.48,45) .. controls (187.87,45) and (189,46.13) .. (189,47.52) -- (189,58.48) .. controls (189,59.87) and (187.87,61) .. (186.48,61) -- (79.52,61) .. controls (78.13,61) and (77,59.87) .. (77,58.48) -- cycle ;
\draw [color={rgb, 255:red, 0; green, 0; blue, 0 }  ,draw opacity=1 ]   (85,61) -- (85,69) ;
\draw  [color={rgb, 255:red, 0; green, 0; blue, 0 }  ,draw opacity=1 ][fill={rgb, 255:red, 255; green, 255; blue, 255 }  ,fill opacity=1 ] (101,29) .. controls (101,24.58) and (104.58,21) .. (109,21) -- (109,21) .. controls (113.42,21) and (117,24.58) .. (117,29) -- (117,29) .. controls (117,33.42) and (113.42,37) .. (109,37) -- (109,37) .. controls (104.58,37) and (101,33.42) .. (101,29) -- cycle ;
\draw [color={rgb, 255:red, 0; green, 0; blue, 0 }  ,draw opacity=1 ]   (109,37) -- (109,45) ;
\draw [color={rgb, 255:red, 0; green, 0; blue, 0 }  ,draw opacity=1 ]   (117,29) -- (125,29) ;
\draw  [color={rgb, 255:red, 0; green, 0; blue, 0 }  ,draw opacity=1 ][fill={rgb, 255:red, 255; green, 255; blue, 255 }  ,fill opacity=1 ] (125,29) .. controls (125,24.58) and (128.58,21) .. (133,21) -- (133,21) .. controls (137.42,21) and (141,24.58) .. (141,29) -- (141,29) .. controls (141,33.42) and (137.42,37) .. (133,37) -- (133,37) .. controls (128.58,37) and (125,33.42) .. (125,29) -- cycle ;
\draw [color={rgb, 255:red, 0; green, 0; blue, 0 }  ,draw opacity=1 ]   (133,37) -- (133,45) ;
\draw [color={rgb, 255:red, 0; green, 0; blue, 0 }  ,draw opacity=1 ]   (141,29) -- (149,29) ;
\draw  [color={rgb, 255:red, 0; green, 0; blue, 0 }  ,draw opacity=1 ][fill={rgb, 255:red, 255; green, 255; blue, 255 }  ,fill opacity=1 ] (149,29) .. controls (149,24.58) and (152.58,21) .. (157,21) -- (157,21) .. controls (161.42,21) and (165,24.58) .. (165,29) -- (165,29) .. controls (165,33.42) and (161.42,37) .. (157,37) -- (157,37) .. controls (152.58,37) and (149,33.42) .. (149,29) -- cycle ;
\draw [color={rgb, 255:red, 0; green, 0; blue, 0 }  ,draw opacity=1 ]   (157,37) -- (157,45) ;
\draw [color={rgb, 255:red, 0; green, 0; blue, 0 }  ,draw opacity=1 ]   (165,29) -- (173,29) ;
\draw  [color={rgb, 255:red, 0; green, 0; blue, 0 }  ,draw opacity=1 ][fill={rgb, 255:red, 255; green, 255; blue, 255 }  ,fill opacity=1 ] (173,29) .. controls (173,24.58) and (176.58,21) .. (181,21) -- (181,21) .. controls (185.42,21) and (189,24.58) .. (189,29) -- (189,29) .. controls (189,33.42) and (185.42,37) .. (181,37) -- (181,37) .. controls (176.58,37) and (173,33.42) .. (173,29) -- cycle ;
\draw [color={rgb, 255:red, 0; green, 0; blue, 0 }  ,draw opacity=1 ]   (181,37) -- (181,45) ;
\draw [color={rgb, 255:red, 0; green, 0; blue, 0 }  ,draw opacity=1 ]   (189,29) -- (197,29) ;
\draw [color={rgb, 255:red, 0; green, 0; blue, 0 }  ,draw opacity=1 ]   (213,29) -- (221,29) ;
\draw  [color={rgb, 255:red, 0; green, 0; blue, 0 }  ,draw opacity=1 ][fill={rgb, 255:red, 255; green, 255; blue, 255 }  ,fill opacity=1 ] (221,29) .. controls (221,24.58) and (224.58,21) .. (229,21) -- (229,21) .. controls (233.42,21) and (237,24.58) .. (237,29) -- (237,29) .. controls (237,33.42) and (233.42,37) .. (229,37) -- (229,37) .. controls (224.58,37) and (221,33.42) .. (221,29) -- cycle ;
\draw [color={rgb, 255:red, 0; green, 0; blue, 0 }  ,draw opacity=1 ]   (229,37) -- (229,45) ;
\draw [color={rgb, 255:red, 0; green, 0; blue, 0 }  ,draw opacity=1 ]   (109,61) -- (109,69) ;
\draw [color={rgb, 255:red, 0; green, 0; blue, 0 }  ,draw opacity=1 ]   (133,61) -- (133,69) ;
\draw [color={rgb, 255:red, 0; green, 0; blue, 0 }  ,draw opacity=1 ]   (157,61) -- (157,69) ;
\draw [color={rgb, 255:red, 0; green, 0; blue, 0 }  ,draw opacity=1 ]   (181,61) -- (181,69) ;
\draw [color={rgb, 255:red, 0; green, 0; blue, 0 }  ,draw opacity=1 ] [dash pattern={on 0.75pt off 0.75pt}]  (53,29) -- (59,29) ;
\draw [color={rgb, 255:red, 0; green, 0; blue, 0 }  ,draw opacity=1 ] [dash pattern={on 0.75pt off 0.75pt}]  (63,29) -- (69,29) ;
\draw [color={rgb, 255:red, 0; green, 0; blue, 0 }  ,draw opacity=1 ] [dash pattern={on 0.75pt off 0.75pt}]  (197,29) -- (203,29) ;
\draw [color={rgb, 255:red, 0; green, 0; blue, 0 }  ,draw opacity=1 ] [dash pattern={on 0.75pt off 0.75pt}]  (207,29) -- (213,29) ;
\draw (13,29) node  [font=\scriptsize]  {$\Omega _{1}$};
\draw (37,29) node  [font=\scriptsize]  {$\Omega _{2}$};
\draw (133,53) node  [font=\footnotesize]  {$\phi ^{\dagger }$};
\draw (133,29) node  [font=\scriptsize]  {$\Omega _{j}$};
\draw (229,29) node  [font=\scriptsize]  {$\Omega _{L}$};
\draw (109,32.8) node [anchor=south] [inner sep=0.75pt]  [font=\scriptsize]  {$\dotsc $};
\draw (157,32.8) node [anchor=south] [inner sep=0.75pt]  [font=\scriptsize]  {$\dotsc $};
\end{tikzpicture}
\ .
\end{equation}
When $l \gg \xi $, convergence in $l$ can be verified numerically. 
In \cref{subsec:gsingle} we provide a self-verification strategy for the creation operator $\hat{\phi}_{j}^{\dagger}$, to ensure that its action is reliable when applied to the DMRG ground state of the large system.

A single-quasiparticle wave packet reads
\begin{equation}
|\Psi \rangle = \hat{\Psi}^\dagger |\Omega_L \rangle 
\ , \quad
\hat{\Psi}^\dagger =\frac{1}{\mathcal{N}}\sum _{j=1}^{L} c_{j}
\hat{\phi }_{j}^{\dagger } \ ,\quad \label{eq:wpcreationop}
\end{equation}
where $c_{j}$ are generic complex coefficients and $\mathcal{N} ={\textstyle \sum _{j}} |c_{j} |^{2}$ is the normalization factor. 
In Appendix \ref{app:wavepacket} we show how to compute an MPO representation of $\hat{\Psi}^\dagger$ given $\hat{\phi }^{\dagger }$ as input.

\subsection{Detection via local measurement}
\label{subsec:detection}

The states obtained by acting on the vacuum with $N$ creation operators supported on well-separated domains
\begin{equation}
|\phi _{j_{1}} ...\phi _{j_{N}} \rangle 
=\hat{\phi }_{j_{1}}^{\dagger } 
...\hat{\phi }_{j_{N}}^{\dagger } 
|\Omega _{L} \rangle 
\end{equation}
span the asymptotic scattering states.
Thus, two wave-packet creation operators $ \hat{\Psi}_\mathcal{A}^\dagger ,\hat{\Psi }_\mathcal{B}^\dagger$ generate a two-wave-packet state $\hat{\Psi }_\mathcal{B}^\dagger \hat{\Psi }_\mathcal{A}^\dagger |\Omega_L \rangle$ if the two supports $\mathcal{A}, \mathcal{B}$ are spatially well separated. 
An input MPS scattering state can be prepared via this protocol, and then time evolved via an MPS time-evolution algorithm, such as the time dependent variational principle (TDVP) \cite{Haegeman2011TimedependentVariationalPrinciple}.
By projecting a time-evolved state onto this basis, one can also extrapolate the $S$-matrix elements of the theory \cite{Rigobello2021EntanglementGeneration1+1D}.

In this section, we provide another detection strategy based on local measurements acting on a finite lattice support $\mathcal{S}_j$ centered at site $j$, with the same extension as $\mathcal{W}$.
We first observe that, being the theory interacting, there is no obvious local dressed-quasiparticle number operator; in particular $\hat \phi^\dagger \hat \phi$ is not informative, being $\hat \phi^\dagger$ unitary.

Given a state $| \Psi \rangle$, the physical information about the region $\mathcal{S}_j$ is contained in the associated reduced density matrix (RDM) $\hat \rho^\Psi_j$:
\begin{equation}
\label{eq:rhopsij}
 \begin{array}{l}
\hat{\rho}_{j}^{\Psi} =\Tr_{\overline{\mathcal{S}_j}} |\Psi \rangle \langle \Psi | \qquad \qquad \qquad \qquad \qquad \\
= \input{tikzpictures/rdmphi} \ .
\end{array}
\end{equation}
To detect a single-particle state in this region, we compare this RDM with the one of the MLWF:
\begin{equation}
\label{eq:rhodetector}
\hat{\rho }^{\phi} =\Tr_{\overline{\mathcal{W}}} 
|\phi _{j_0} \rangle \langle \phi _{j_0} 
|=\tikzset{every picture/.style={line width=0.75pt}} 
\begin{tikzpicture}[x=0.75pt,y=0.75pt,yscale=-1,xscale=1, baseline=(XXXX.south) ]
\path (0,84);\path (142.609375,0);\draw    ($(current bounding box.center)+(0,0.3em)$) node [anchor=south] (XXXX) {};
\draw  [color={rgb, 255:red, 0; green, 0; blue, 0 }  ,draw opacity=1 ][fill={rgb, 255:red, 255; green, 255; blue, 255 }  ,fill opacity=1 ] (5,51.83) .. controls (5,48.06) and (8.06,45) .. (11.83,45) -- (128.17,45) .. controls (131.94,45) and (135,48.06) .. (135,51.83) -- (135,53.17) .. controls (135,56.94) and (131.94,60) .. (128.17,60) -- (11.83,60) .. controls (8.06,60) and (5,56.94) .. (5,53.17) -- cycle ;
\draw [color={rgb, 255:red, 0; green, 0; blue, 0 }  ,draw opacity=1 ]   (70,60) -- (70,70) ;
\draw [color={rgb, 255:red, 0; green, 0; blue, 0 }  ,draw opacity=1 ]   (80,60) -- (80,70) ;
\draw [color={rgb, 255:red, 0; green, 0; blue, 0 }  ,draw opacity=1 ]   (60,60) -- (60,70) ;
\draw [color={rgb, 255:red, 0; green, 0; blue, 0 }  ,draw opacity=1 ]   (50,60) -- (50,70) ;
\draw [color={rgb, 255:red, 0; green, 0; blue, 0 }  ,draw opacity=1 ]   (90,60) -- (90,70) ;
\draw  [color={rgb, 255:red, 0; green, 0; blue, 0 }  ,draw opacity=1 ][fill={rgb, 255:red, 255; green, 255; blue, 255 }  ,fill opacity=1 ] (5,31.83) .. controls (5,28.06) and (8.06,25) .. (11.83,25) -- (128.17,25) .. controls (131.94,25) and (135,28.06) .. (135,31.83) -- (135,33.17) .. controls (135,36.94) and (131.94,40) .. (128.17,40) -- (11.83,40) .. controls (8.06,40) and (5,36.94) .. (5,33.17) -- cycle ;
\draw    (20,25) .. controls (8.66,3.37) and (8.99,79.87) .. (20,60) ;
\draw    (10,25) .. controls (-1.34,3.37) and (-1.01,79.87) .. (10,60) ;
\draw    (40,25) .. controls (28.66,3.37) and (28.99,79.87) .. (40,60) ;
\draw    (130,25) .. controls (141.34,3.37) and (141.01,79.87) .. (130,60) ;
\draw    (120,25) .. controls (131.34,3.37) and (131.01,79.87) .. (120,60) ;
\draw    (100,25) .. controls (111.34,3.37) and (111.01,79.87) .. (100,60) ;
\draw [color={rgb, 255:red, 0; green, 0; blue, 0 }  ,draw opacity=1 ]   (70,15) -- (70,25) ;
\draw [color={rgb, 255:red, 0; green, 0; blue, 0 }  ,draw opacity=1 ]   (80,15) -- (80,25) ;
\draw [color={rgb, 255:red, 0; green, 0; blue, 0 }  ,draw opacity=1 ]   (60,15) -- (60,25) ;
\draw [color={rgb, 255:red, 0; green, 0; blue, 0 }  ,draw opacity=1 ]   (50,15) -- (50,25) ;
\draw [color={rgb, 255:red, 0; green, 0; blue, 0 }  ,draw opacity=1 ]   (90,15) -- (90,25) ;
\draw (69,76) node  [font=\scriptsize]  {$\mathcal{W}$};
\draw (70,52.5) node  [font=\footnotesize]  {$\phi _{j_0}$};
\draw (70,32.5) node  [font=\footnotesize]  {$\phi _{j_0}$};
\end{tikzpicture}
\ .\ \
\end{equation}
To detect a single-particle $\phi$ in the region $\mathcal{S}_j$, we need to compute the fidelity between \eqref{eq:rhopsij} and \eqref{eq:rhodetector}. 
A reliable metric for the fidelity between two mixed states $\hat \rho_1, \hat \rho_2$ is the Uhlmann fidelity \cite{uhlmannTransitionProbabilityState1976}:
\begin{equation}
F[\hat\rho_1, \hat\rho_2] =
\left(\Tr\sqrt{\sqrt{\hat\rho_1},
\hat\rho_2\,\sqrt{\hat\rho_1}}
\right)^{2}.
\end{equation}
This fidelity is computable numerically (since $\mathcal{S}_j$ is a small support), but is not an observable and therefore not obtainable from local measurements in a quantum simulator.

To provide a state discrimination strategy via local measurement, we compute the Hilbert-Schmidt (HS) overlap instead, which corresponds to a quasi-local observable with support $\mathcal{S}$:
\begin{equation}
\begin{array}{l}
\Tr\left[\hat{\rho }^{\phi }\hat{\rho }_{j}^{\Psi }\right]
= \langle \Psi | \hat \rho^\phi_j | \Psi \rangle
\qquad \qquad \qquad 
\\
= \input{tikzpictures/detection_diagram} \, .
\end{array}
\end{equation}
The HS overlap is not a valid measure of fidelity for mixed quantum states. 
However, one can prove \cite{miszczakSubSuperfidelityBounds2008} the following fidelity bound
\begin{equation}
\label{eq:superfidelity}
\Tr[\hat{\rho }_{1}\hat{\rho }_{2}] 
\leqslant F[\hat{\rho }_{1} ,\hat{\rho }_{2}] 
\leqslant \Tr[\hat{\rho }_{1}\hat{\rho }_{2}] 
+\Delta [\hat{\rho }_{1} ,\hat{\rho }_{2}] \ ,
\end{equation}
where $\Delta$ is a correction which is a function of the purities of $\hat \rho_1$ and $\hat \rho_2$:
\begin{equation}
\label{eq:mixednesscorrection}
\Delta [\hat{\rho }_{1} ,\hat{\rho }_{2}] 
=\sqrt{1-\Tr\left[\hat{\rho }_{1}^{2}\right]}
\sqrt{1-\Tr\left[\hat{\rho }_{2}^{2}\right]} \ .
\end{equation}
If the HS overlap is high, from the lower bound of \cref{eq:superfidelity} also $F$ is high, and we can infer the presence of a $\phi$ quasiparticle in the region $\mathcal{S}_j$ for the state $| \Psi \rangle$.
If $\hat \rho^\phi$ or $\hat \rho^\Psi_j$ is close to a pure state (e.g., for low-entangled vacua), then $\Delta \sim 0$ and $F$ is close to the HS overlap. 
In particular, if one state is pure (e.g., when the ground state is a product state), $F$ is exactly the HS overlap.

One can detect multiple quasiparticle species $\phi_1, \phi_2$ by applying the same strategy, computing the different RDMs from different MLWFs.
However, the problem of detecting unknown quasiparticles, such as resonances, is still open because $\mathbb{I} -\hat{\rho }^{\phi _{1}} 
-\hat{\rho }^{\phi _{2}} - \ldots$ includes also the contribution of the vacuum. 
The proposed method to discriminate resonances, not based on measurement of observables, is described in Appendix \ref{app:resonances}.

\section{The models}

\label{sec:model}

We consider a Hamiltonian lattice gauge theory with gauge group $G$ in the Kogut--Susskind formulation \cite{Kogut1975HamiltonianFormulationWilsons, Kogut1979IntroductionLatticeGauge}. 
The general theoretical framework is summarized in Appendix \ref{app:hamiltonian}; here we introduce the ladder geometry and the specific model studied in our simulations.

\subsection{LGT on a ladder geometry}

\label{subsec:ladder}

\begin{figure}[t]
    \centering
    \hspace*{-1mm}\includegraphics[scale=0.95]{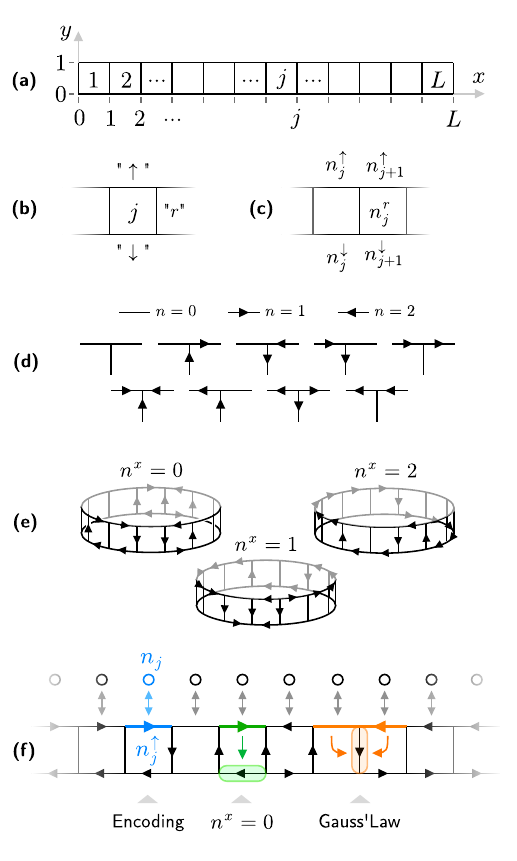}
    \caption{\letterimagecaption{a} OBC ladder with $L$ plaquettes and $L+1$ rungs ($L$ in PBC); lattice spacing $a=1$. 
    \letterimagecaption{b} Notation: $\uparrow$, $\downarrow$, $r$ for rails and rungs. \letterimagecaption{c} Link DoFs $n_j^{\uparrow}$, $n_j^{\downarrow}$, 
    $n_j^r$, $n_{j-1}^r$ at plaquette $j$. 
    \letterimagecaption{d} The nine $\top$-junction configurations compatible with 
    Gauss' law ($\Z{3}$ or $\SU{3}_1$). 
    \letterimagecaption{e} Examples of states in the $n$ basis for the three polarization sectors $n^x = 0,1,2$.
    \letterimagecaption{f} In the $n^x = 0$ sector, both truncations map to a qutrit chain with local DoFs $n_j^\uparrow$.}
    \label{fig:ladder}
\end{figure}

In $1+1$ dimensions, gauge bosons do not propagate due to the lack of a transverse polarization \cite{weinbergQuantumTheoryFields1996,colemanMoreMassiveSchwinger1976}. 
The minimal lattice that hosts non-trivial gauge dynamics is therefore a ladder of length $L$ (\cref{fig:ladder}\letterimagetext{a}): the rungs of the ladder play the role of a synthetic dimension, providing the transverse degree of freedom (DoF) needed to sustain propagating gauge excitations, analogous to a transverse electric field polarization in QED. 
We use the conventions in 
\cref{fig:ladder}\letterimagetext{b,}\letterimagetext{c} to name links 
and plaquettes.

Let $T^a$ be the (Hermitian) generators of the Lie algebra of the gauge group $G$, and $\hat{L}_{\ell}^a, \hat{R}_{\ell}^a$ be the left and right gauge transformation operators on a link $\ell$. The electric field operator $\hat{E}_{\ell }^{2}$ is defined as the Casimir operator
\begin{equation}
\hat{E}_{\ell}^{2} =\sum_a (\hat{L}_{\ell }^a)^2 =\sum_a (\hat{R}_{\ell}^a)^2 \ ,
\end{equation}
while the parallel transporter can be defined via commutation relations
\begin{equation}
[\hat{L}_{\ell}^a ,\hat{U}_{\ell^\prime }] =-\delta _{\ell \ell^\prime } \thinspace T^a\hat{U}_{\ell } \ ,\ \ [\hat{R}_{\ell}^a ,\hat{U}_{\ell^\prime }] =+\delta _{\ell \ell^\prime } \thinspace \hat{U}_{\ell } T^a \ .
\end{equation}
The Hamiltonian of the model takes the form
\begin{equation}
\hat{H} = \lambda\,\hat{H}_E + (1-\lambda)\,
\hat{H}_B \, , \label{eq:ladderham}
\end{equation}
where $\lambda(g) = g^4 / (1 + g^4)$ is a monotonically 
increasing function of the gauge coupling $g$ with 
$\lambda \to 0$ for $g^2 \to 0$ (weak coupling, 
magnetic dominance) and $\lambda \to 1$ for 
$g^2 \to \infty$ (strong coupling, electric dominance). The electric and magnetic contributions are (see \cref{fig:ladder}\letterimagetext{b} for the link and plaquette naming convention)
\begin{equation}
\hat{H}_{E} =\sum _{j}\hat{E}_{j}^{r2} 
+\hat{E}_{j}^{\uparrow 2} +\hat{E}_{j}^{\downarrow 2} 
\ , \ \
\hat{H}_{B} = -\sum _{j}\hat{U}_{j} 
+\hat{U}_{j}^{\dagger} \ ,
\end{equation}
where $\hat{U}_j = \Tr[\hat{U}_j^{\downarrow}
\hat{U}_j^r \hat{U}_j^{\uparrow\dagger}
\hat{U}_{j-1}^{r\dagger}]$ is the $j$-th plaquette 
operator.
The local Hamiltonian $\hat{h}_j$ (see \cref{subsec:spectral}) is computed with respect to a plaquette $j$, distributing the electric energy symmetrically on the neighboring rungs.

We truncate the link Hilbert space to a finite 
dimension. For $G = \U{1}$ we use a 
$\Z{3} \subset \U{1}$ subgroup approximation 
\cite{magnificoRealTimeDynamics2020,
Pradhan2024DiscreteAbelianLattice}; 
for $G = \SU{3}$ we restrict to 
the irreducible representations (irreps) 
$\boldsymbol{1}, \boldsymbol{3}, \overline{\boldsymbol{3}}$.
This amounts to an electric energy density cutoff 
that retains only representations up to the first Casimir
level, a.k.a.\ hardcore-gluon approximation
\cite{Rigobello2023Hadrons1plus1DHamiltonian, Cataldi20232+1DSU2YangMills}.
Mathematically, this corresponds to a level-1 Chern–Simons 
theory $\SU{3}_1$, which is a pointed theory whose fusion 
ring is isomorphic to $\Z{3}$
\cite{rowellClassificationModularTensor2009,
bruillardClassificationSupermodularCategories2017a}.
The $\Z{3}$ and $\SU{3}_1$ fusion rules read
\begin{align}
\eta_n \otimes \eta_m &= \eta_{n+m\
\text{mod}\thinspace 3} \, , \label{eq:z3fusionrule} \\
\mathbf{j}_n \otimes \mathbf{j}_m &= \mathbf{j}_{n+m\ 
\text{mod}\thinspace 3} \oplus \text{truncated}
\, ,
\label{eq:su3fusionrule}
\end{align}
where $\eta_n$ and $\mathbf{j}_n$ are respectively $\Z{3}$ and $\SU{3}_1$ irreps, with 
$\mathbf{j}_0 = \boldsymbol{1}$, 
$\mathbf{j}_1 = \boldsymbol{3}$,
$\mathbf{j}_2 = \overline{\boldsymbol{3}}$, and 
``truncated'' denotes the higher 
representations discarded by the truncation. 

The $\SU{3}_1$ Hilbert space has dimension 19 (see \cite{Rigobello2023Hadrons1plus1DHamiltonian}).
However, by focusing on a single lattice site and fixing the representations attached to it, Gauss’ law uniquely determines a single gauge-invariant singlet state. This construction is known as the dressed site formalism (see Appendix \ref{app:dressed_site_formalism}).
Consequently, gauge-invariant states are fully characterized by the irreps assigned to the links. In both Abelian and non-Abelian settings, link states can therefore be labeled by $n \in \{0,1,2\}$. The gauge-invariant $\top$-junction states satisfy
\begin{equation}
n_j^\uparrow - n_{j+1}^\uparrow + n_j^r = 0 \mod 3 ,
\end{equation}
giving the nine $\top$-junction configurations in \cref{fig:ladder}\letterimagetext{d}.

\subsection{The ladder-clock duality}
\label{subsec:duality}

In this section, we define the qutrit chain encoding of the above ladder models used in numerical simulations. 
The duality between $\Z{N}$ LGTs on a ladder and quantum clock models has already been studied in \cite{Pradhan2024DiscreteAbelianLattice}. 
We provide an alternative derivation of the same duality, which also generalizes to $\SU{3}_1$.
Here we summarize the main results; details are provided in Appendix \ref{app:operators}.

In the electric basis, the operators $\hat{n}_{\ell }$ are diagonal. 
Let $|n_{\ell } \rangle $ denote the electric field configuration where the local state on each link $\ell $ is labeled by $n_{\ell }$. 
This basis spans a (total, non gauge-invariant) Hilbert space $\mathcal{H}$ of dimension $\dim\mathcal{H} =3^{\#\ell }$ where $\#\ell $ is the number of links. 
However, the residual Gauss' Law constraints further reduces the space to a smaller one $\mathcal{H}_{\text{phys}}$,
by enforcing
\begin{equation}
\begin{cases}
\label{eq:gaussconstr}
n_{j}^{\uparrow } +n_{j}^{r} =n_{j+1}^{\uparrow } & \\
n_{j}^{\downarrow } -n_{j}^{r} =n_{j+1}^{\downarrow } & 
\end{cases} \mod 3 \ .
\end{equation}
We define the longitudinal polarization of the $j$-th plaquette as
\begin{equation}
n_{j}^{x} = n_{j}^{\uparrow } + n_{j}^{\downarrow} \mod 3 \, ,
\end{equation}
and from the constraint \eqref{eq:gaussconstr} it follows immediately, for all the basis elements:
\begin{equation}
n_{j}^{x} =n_{j+1}^{x} \quad \forall j \, .
\end{equation}
By induction, $n_{j}^{x}$ is the same for every plaquette $j$, hence the gauge invariant Hilbert space splits in three longitudinal polarization sectors, labeled by $n^{x}$ (see \cref{fig:ladder}\letterimagetext{e}):
\begin{gather}
\label{eq:polarizsectors}
\mathcal{H}_{\text{phys}} = \mathcal{H}_{0} \oplus \mathcal{H}_{1} \oplus \mathcal{H}_{2} \ ,\\
\mathcal{H}_{n^{x}} \equiv \text{span}\left\{|n_{\ell } \rangle \ |\ n_{j}^{x} =n^{x}\;\forall j\right\} \ .
\end{gather}
These sectors are also dynamically decoupled since,
\begin{equation}
[ \hat{n}_{j}^{x} ,\hat{H} ] =0 \quad \forall j\ ;
\end{equation}
indeed,
\begin{itemize}
\item the electric field operator $\hat{E}_{\ell }^{2}$ is diagonal in the $|n_{\ell } \rangle $ basis, hence it commutes with $\hat{n}_{\ell }$, and so with $\hat{n}_{j}^{x}$;
\item the action of the plaquette operator $\hat{U}_{j}$ on every element of the basis $|n_{\ell } \rangle $ does not change the longitudinal polarization in the plaquette $j$, hence $\hat{U}_{j}$ commutes with $\hat{n}_{j}^{x}$.
\end{itemize}

In general, due to gauge constraints, the space of an LGT is not a local tensor-product Hilbert space. 
However, we notice that, choosing a state for all the plaquette top links $( n_{j}^{\uparrow } )_{j=1}^{L}$ then (see \cref{fig:ladder}\letterimagetext{f}):
\begin{itemize}
\item the bottom link states are derived by the longitudinal polarization conservation $n_{j}^{\downarrow } =n_{j}^{x} -n_{j}^{\uparrow }$;
\item the rung states are derived by the Gauss' Law constraint $n_{j}^{r} =n_{j+1}^{\uparrow } -n_{j}^{\uparrow }$.
\end{itemize}

Hence, the independent DoFs are only $n^\uparrow_j$, so each of the sectors in \cref{eq:polarizsectors} is injectively mapped to the Hilbert space of a qutrit chain with $L$ sites
\begin{equation}
\mathcal{H}_{n^{x}} \ \rightarrow \ \left(\mathbb{C}^{3}\right)^{\otimes L} \ .
\end{equation}

To map the Hamiltonian in \eqref{eq:ladderham} onto this spin chain, we need to compute the electric and magnetic local operators in the eigenbasis of $\hat{n}$. 
The electric operator $\hat{E}_{\ell}^{2}$ is diagonal:
\begin{equation}
\label{eq:electricfielddiag}
\hat{E}_{\ell }^{2} ={\textstyle \begin{pmatrix}
0 &  & \\
 & C_{2} & \\
 &  & C_{2}
\end{pmatrix} \ ,}
\end{equation}
with
\begin{itemize}
\item $ C_{2} =27/4\pi ^{2}$ for $\Z{3}$, obtained from a discretized version of the Laplacian operator of $\U{1}$ 
\cite{zoharQuantumSimulationsGauge2013,Ercolessi2018PhaseTransitionsZ_ntextbackslash};
\item $C_{2} = 4/3$ for $ \SU{3}_{1}$, namely the Casimir of 
the (anti-)fundamental representation
\cite{peskinIntroductionQuantumField2018}.
\end{itemize}

Recalling the definition of the three-state clock model, conjugate variables
\begin{equation}
\label{eq:clockvariables}
\hat{\sigma } =\begin{pmatrix}
1 &  & \\
 & \eta  & \\
 &  & \eta ^{2}
\end{pmatrix} \ ,\quad \hat{\tau } =\begin{pmatrix}
 &  & 1\\
1 &  & \\
 & 1 & 
\end{pmatrix} \ ,
\end{equation}
with $ \hat{\sigma }\hat{\tau } 
=\eta \hat{\tau }\hat{\sigma }$ and 
$ \eta =e^{2\pi i/3}$, one obtains the following expression 
for the electric energy term (see Appendix \ref{app:electricmapping} for details)
\begin{equation}
\label{eq:HE_mapping_potts}
\hat{H}_{E} = -\frac{C_{2}}{3} \sum_{j}
\bigl(\hat{\sigma}_{j}^{\dagger}\hat{\sigma}_{j+1}
+ 2\,\hat{\sigma}_{j} + \text{H.c.}\bigr) \, .
\end{equation}
This is the interaction term of the three-state clock model with an additional longitudinal field that breaks integrability.
We stress that Eq. \eqref{eq:HE_mapping_potts} is valid for both $\Z{3}$ and $\SU{3}_1$ theories.
Note that for $\lambda =1$ ($ g\rightarrow \infty $) we have $ \hat{H} =\hat{H}_{E}$ and the two theories have the same Hamiltonian up to a rescaling.

For the magnetic term $ \hat{H}_{B}$, it is sufficient to map the operator $ \hat{U}_j$. We will provide a summary of the results here (see Appendix \ref{app:operators} for details):
\begin{itemize}
\item for $\Z{3}$, the plaquette term $ 
\hat{U}_j$ is mapped to a one-body term, 
which corresponds to the transverse field term in 
the three-state clock model:
\begin{equation}
\hat{U}_j = \hat{\tau}_{j}^{\dagger} \qquad
\hat{U}_j^{\dagger} = \hat{\tau}_{j} \ ;
\end{equation}
\item for $ \SU{3}_{1}$, the operator $\hat{U}_j$ is a three-body term whose matrix elements are computed from the Clebsch--Gordan coefficients of $\SU{3}$ (see \cref{fig:plaquettemels}).
\end{itemize}

\subsection{Quasiparticle classification at 
$g^2 \rightarrow \infty$ for $n^x = 0$}

\label{subsec:classification}

The strong-coupling limit classification can be done at $\lambda = 1$, where the eigenstates are product states in the electric basis, and the quasiparticle quantum numbers can be read off analytically.
In this regime, the spectrum bands are flat (no propagation).
Hence, this limit provides a natural starting point for classifying the low-energy glueball spectrum: each state can be identified unambiguously by the energy and other quantum numbers, and the labels remain valid at finite coupling as long as band crossings do not occur (adiabatic continuity).

Particles are classified by the irreps of the theory's symmetry group. In particular, the Bloch states transform under irreps of $\hat{H}$ and $\hat{T}$, which generate the time and (discrete) space translations. 
The residual symmetries in the ladder are generated by (see \cref{fig:ladder} for reference):
\begin{itemize}
\item the spatial reflection in the extensive $x$-dimension $x\mapsto L - x$, which corresponds to the parity operator $\hat{P}$;
\item the spatial reflection in the synthetic $y$-dimension $y\mapsto 1 - y$, which in the $n^{x} =0$ polarization sector is also identified by the charge conjugation operator $\hat C$, since $\hat{n}_{j}^{\uparrow } =-\hat{n}_{j}^{\downarrow }$.
\end{itemize}
The operator $ \hat{C}$ commutes with $ \hat{T}$, whereas $ \hat{P}$ does not. 
Nevertheless, one can identify the parity of the band as that defined for the $k = 0$ state; in this sector, $\hat T$ and $\hat P$ commute and both quantum numbers are well-defined.

The residual symmetry group 
for each quasiparticle is then:
\begin{equation}
G_{\text{res}} =\Z{2}^{( P)} \times 
\Z{2}^{( C)} \ .
\end{equation}
This group is Abelian, hence there are four 1-dimensional irreps, which we can label by $(\hat{P} ,\hat{C} )$ or, in analogy with the standard glueball nomenclature, with the $J^{PC}$ notation (\cref{tab:sectors} lists the symmetry sectors). 
Notice that there is no spin analogue in $ ( 1+1)$D (hence $ J=0$) since any spacetime rotation is recovered in the continuum limit in one spatial dimension.

\begin{table}[]
    \begin{center}
\begin{tabular}{ccccc}
\toprule 
 $ P$ & $ C$ & $ J^{PC}$ & Sector name & Symbol \\
\midrule 
 $ +$ & $ +$ & $ 0^{++}$ & Scalar & $\textcolor[rgb]{0,0.65,1}{\bigstar}$ \\
$ -$ & $ +$ & $ 0^{-+}$ & Pseudoscalar & $\textcolor[rgb]{0,0.65,1}{\blacktriangle}$ \\
$ +$ & $ -$ & $ 0^{+-}$ & Exotic scalar & $\textcolor[rgb]{0.25,0.85,0}{\bigstar}$ \\
$ -$ & $ -$ & $ 0^{--}$ & Exotic pseudoscalar & $\textcolor[rgb]{0.25,0.85,0}{\blacktriangle}$ \\
 \bottomrule
\end{tabular}
\end{center}

    \caption{The four ladder glueball sectors. 
    The notation $ J^{PC}$ and the naming of the sector comes from the standard notation used in high-energy physics for the particles in space-time dimension (3+1). 
    The last column shows the symbols used to denote each sector in the excitation spectrum (\cref{fig:disprel}).}
    \label{tab:sectors}
\end{table}

At $\lambda = 1$, only the electric term dominates, and the plaquette term 
does not contribute to the energy shift. In this regime, the $\Z{3}$ and $\SU{3}_1$ Hamiltonians are equivalent (up to a rescaling) and thus share analogous excitations.

The simplest single-quasiparticle configurations are the two single-plaquette glueballs, which are also the two lightest states (see \cref{tab:sectors,tab:classification}):
\begin{equation}
|0_{1}^{++} \rangle =\frac{|\tikzset{every picture/.style={line width=0.75pt}} 
\begin{tikzpicture}[x=0.75pt,y=0.75pt,yscale=-1,xscale=1, baseline=(XXXX.south) ]
\path (0,19);\path (19.5,0);\draw    ($(current bounding box.center)+(0,0.3em)$) node [anchor=south] (XXXX) {};
\draw    (3,3) -- (15,3) ;
\draw [shift={(11.6,3)}, rotate = 180] [fill={rgb, 255:red, 0; green, 0; blue, 0 }  ][line width=0.08]  [draw opacity=0] (5.36,-2.57) -- (0,0) -- (5.36,2.57) -- cycle    ;
\draw    (3,15) -- (3,3) ;
\draw [shift={(3,6.4)}, rotate = 90] [fill={rgb, 255:red, 0; green, 0; blue, 0 }  ][line width=0.08]  [draw opacity=0] (5.36,-2.57) -- (0,0) -- (5.36,2.57) -- cycle    ;
\draw    (15,3) -- (15,15) ;
\draw [shift={(15,11.6)}, rotate = 270] [fill={rgb, 255:red, 0; green, 0; blue, 0 }  ][line width=0.08]  [draw opacity=0] (5.36,-2.57) -- (0,0) -- (5.36,2.57) -- cycle    ;
\draw    (15,15) -- (3,15) ;
\draw [shift={(6.4,15)}, rotate = 360] [fill={rgb, 255:red, 0; green, 0; blue, 0 }  ][line width=0.08]  [draw opacity=0] (5.36,-2.57) -- (0,0) -- (5.36,2.57) -- cycle    ;
\end{tikzpicture}
\rangle +|\tikzset{every picture/.style={line width=0.75pt}} 
\begin{tikzpicture}[x=0.75pt,y=0.75pt,yscale=-1,xscale=1, baseline=(XXXX.south) ]
\path (0,19);\path (19.5,0);\draw    ($(current bounding box.center)+(0,0.3em)$) node [anchor=south] (XXXX) {};
\draw    (3,15) -- (15,15) ;
\draw [shift={(11.6,15)}, rotate = 180] [fill={rgb, 255:red, 0; green, 0; blue, 0 }  ][line width=0.08]  [draw opacity=0] (5.36,-2.57) -- (0,0) -- (5.36,2.57) -- cycle    ;
\draw    (3,3) -- (3,15) ;
\draw [shift={(3,11.6)}, rotate = 270] [fill={rgb, 255:red, 0; green, 0; blue, 0 }  ][line width=0.08]  [draw opacity=0] (5.36,-2.57) -- (0,0) -- (5.36,2.57) -- cycle    ;
\draw    (15,15) -- (15,3) ;
\draw [shift={(15,6.4)}, rotate = 90] [fill={rgb, 255:red, 0; green, 0; blue, 0 }  ][line width=0.08]  [draw opacity=0] (5.36,-2.57) -- (0,0) -- (5.36,2.57) -- cycle    ;
\draw    (15,3) -- (3,3) ;
\draw [shift={(6.4,3)}, rotate = 360] [fill={rgb, 255:red, 0; green, 0; blue, 0 }  ][line width=0.08]  [draw opacity=0] (5.36,-2.57) -- (0,0) -- (5.36,2.57) -- cycle    ;
\end{tikzpicture}
\rangle }{\sqrt{2}} ,\quad |0_{1}^{--} \rangle =\frac{|\tikzset{every picture/.style={line width=0.75pt}} 
\begin{tikzpicture}[x=0.75pt,y=0.75pt,yscale=-1,xscale=1, baseline=(XXXX.south) ]
\path (0,19);\path (19.5,0);\draw    ($(current bounding box.center)+(0,0.3em)$) node [anchor=south] (XXXX) {};
\draw    (3,3) -- (15,3) ;
\draw [shift={(11.6,3)}, rotate = 180] [fill={rgb, 255:red, 0; green, 0; blue, 0 }  ][line width=0.08]  [draw opacity=0] (5.36,-2.57) -- (0,0) -- (5.36,2.57) -- cycle    ;
\draw    (3,15) -- (3,3) ;
\draw [shift={(3,6.4)}, rotate = 90] [fill={rgb, 255:red, 0; green, 0; blue, 0 }  ][line width=0.08]  [draw opacity=0] (5.36,-2.57) -- (0,0) -- (5.36,2.57) -- cycle    ;
\draw    (15,3) -- (15,15) ;
\draw [shift={(15,11.6)}, rotate = 270] [fill={rgb, 255:red, 0; green, 0; blue, 0 }  ][line width=0.08]  [draw opacity=0] (5.36,-2.57) -- (0,0) -- (5.36,2.57) -- cycle    ;
\draw    (15,15) -- (3,15) ;
\draw [shift={(6.4,15)}, rotate = 360] [fill={rgb, 255:red, 0; green, 0; blue, 0 }  ][line width=0.08]  [draw opacity=0] (5.36,-2.57) -- (0,0) -- (5.36,2.57) -- cycle    ;
\end{tikzpicture}
\rangle -|\tikzset{every picture/.style={line width=0.75pt}} 
\begin{tikzpicture}[x=0.75pt,y=0.75pt,yscale=-1,xscale=1, baseline=(XXXX.south) ]
\path (0,19);\path (19.5,0);\draw    ($(current bounding box.center)+(0,0.3em)$) node [anchor=south] (XXXX) {};
\draw    (3,15) -- (15,15) ;
\draw [shift={(11.6,15)}, rotate = 180] [fill={rgb, 255:red, 0; green, 0; blue, 0 }  ][line width=0.08]  [draw opacity=0] (5.36,-2.57) -- (0,0) -- (5.36,2.57) -- cycle    ;
\draw    (3,3) -- (3,15) ;
\draw [shift={(3,11.6)}, rotate = 270] [fill={rgb, 255:red, 0; green, 0; blue, 0 }  ][line width=0.08]  [draw opacity=0] (5.36,-2.57) -- (0,0) -- (5.36,2.57) -- cycle    ;
\draw    (15,15) -- (15,3) ;
\draw [shift={(15,6.4)}, rotate = 90] [fill={rgb, 255:red, 0; green, 0; blue, 0 }  ][line width=0.08]  [draw opacity=0] (5.36,-2.57) -- (0,0) -- (5.36,2.57) -- cycle    ;
\draw    (15,3) -- (3,3) ;
\draw [shift={(6.4,3)}, rotate = 360] [fill={rgb, 255:red, 0; green, 0; blue, 0 }  ][line width=0.08]  [draw opacity=0] (5.36,-2.57) -- (0,0) -- (5.36,2.57) -- cycle    ;
\end{tikzpicture}
\rangle }{\sqrt{2}} \ .
\end{equation}
Under both $ \hat{P}$ and $ \hat{C}$, the first is invariant, while the second flips sign.
All the other quasiparticles can be classified similarly. 
\cref{tab:classification} summarizes the classification of the first 12 lowest-energy quasiparticle glueball excitations for $\lambda =1$.

\begin{table}[]
    \input{tabulars/classification}
    \caption{Classification of the first 12 lowest-energy single-quasiparticle glueball states at $ \lambda =1\ \left( g\rightarrow \infty \right)$. 
    Since the bands are flat at $\lambda = 1$, the energy does not depend on $k$. The excitation basis is defined up to translations of the chain. 
    The quasiparticles are labeled by sectors $J^{PC}_n$, while $n$ is a band index to distinguish quasiparticles in the same sector but with different energies.}
    \label{tab:classification}
\end{table}

\begin{figure}[t]
    \centering
    \includegraphics[scale=1.07]{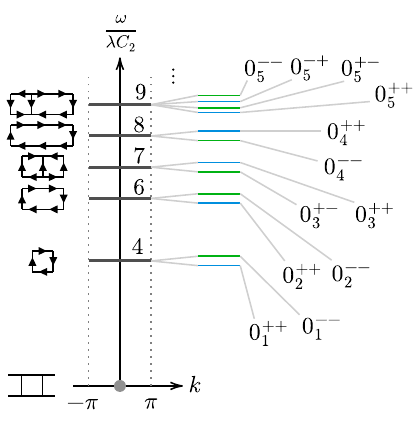}
    \caption{A schematic representation of the low-energy glueball spectrum (grey lines) at $\lambda =1$ $ \left( g\rightarrow \infty \right)$. 
    The spectrum is computed as the electric energy in units of Casimir $ H_{E} /C_{2}$ of the first excitations above the ground state. 
    On the right, a schematic representation of the bands when $ 0< 1-\lambda \ll 1$ ($ g< \infty $ but large) is shown; the degeneracy resolution can be computed using second-order perturbation theory. 
    The blue and green bands belong to the $ C=+1$ and $ C=-1$ sectors, respectively.
    In the $J^{PC}$ glueball notation, $J=0$ because in $(1+1)$D there are no rotational DoFs; hence all states carry the same trivial angular momentum label, and glueball species are distinguished solely by their $P$ and $C$ quantum numbers: $0_j^{\pm,\pm}$, where $j$ indexes the excitation level within each sector.}
    \label{fig:levels}
\end{figure}

The flat band structure is modified by switching on interactions; with $ \lambda < 1$ (finite $ g$), the flat band splits into different bands.
For large $\lambda \sim 1$, the band shape can be estimated from a second-order perturbation expansion in $1/g^4$.
\cref{fig:levels} shows a schematic representation of the first energy levels at $ \lambda =1$, with the splitting into bands, each labelled according to the $ J^{PC}$ convention.

\section{Results}
\label{sec:results}

In this section, we provide numerical results obtained by applying the method introduced in \cref{sec:method} to the models of \cref{sec:model}.

\subsection{Glueball spectra and localization}
\label{subsec:gdisprel}

\begin{figure}[t]
    \centering
    \hspace*{-5mm}\includegraphics{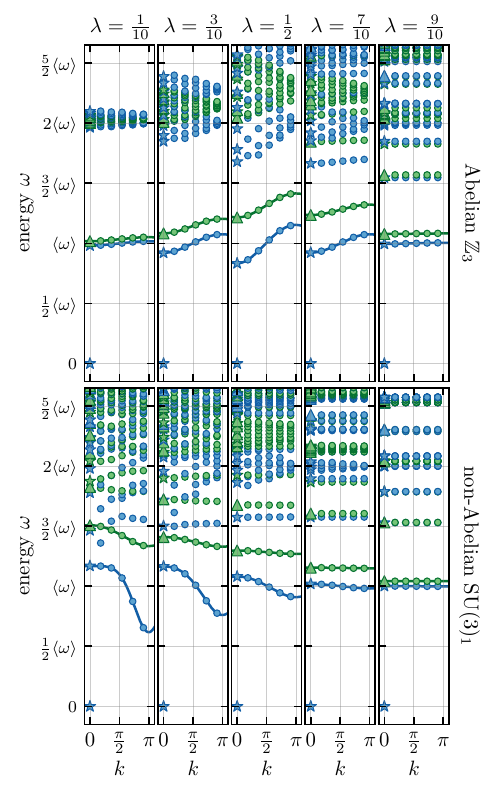}
    \caption{First levels of the excitation spectrum, for $\Z{3}$ (first row) and $\SU{3}_1$ (second row), and increasing couplings $\lambda $ 
    (left to right). In blue ($\textcolor[rgb]{0,0.65,1}{\bullet}$) 
    for the $C=+1$ sector and green 
    ($\textcolor[rgb]{0.25,0.85,0}{\bullet}$) 
    for the $C=-1$ sector. At $k=0$, the 
    parity sector is distinguished by a star shape 
    ($\bigstar$) for $P = +1$ and 
    triangle shape ($\blacktriangle$) for 
    $P = -1$ (see also \cref{tab:sectors}). The continuous line for the first 
    and second band is obtained via Fourier interpolation.}
    \label{fig:disprel}
\end{figure}

\cref{fig:disprel} shows the excitation spectrum of the glueball for both the $\Z{3}$ and $\SU{3}_1$ cases. 
The ground-state energies are shifted to zero, while the energy unit is the first band centroid
\begin{equation}
\label{eq:centroid}
\langle\omega\rangle = \frac{1}{l} \sum_{|k\rangle \in 0^{++}_1} \langle k| \hat H | k \rangle \, .
\end{equation}

In the $\lambda = 0.9$ ($g^2 \gg 1$ regime), we observe that the two spectra are approaching the same spectrum, as expected from the analysis at \cref{subsec:duality}. 
For $\lambda = 0.5$ ($g^2 = 1$), the bands are opened but still cosine-shaped: the effective hopping term is still nearest-neighbor at lowest order in perturbation theory. 
At $\lambda = 0.1$ we approach the $g^2 \sim 0$ limit. 
For $\Z{3}$, the plaquette term is single-body and the Bloch states are separable (no hopping term) thus the bands are flat and there is no propagation. 
By contrast, for $\SU{3}_1$, the three-body plaquette operator makes the theory interacting even in the absence of the electric field term.

\begin{figure}[t]
    \centering
    \hspace*{-5mm}\includegraphics{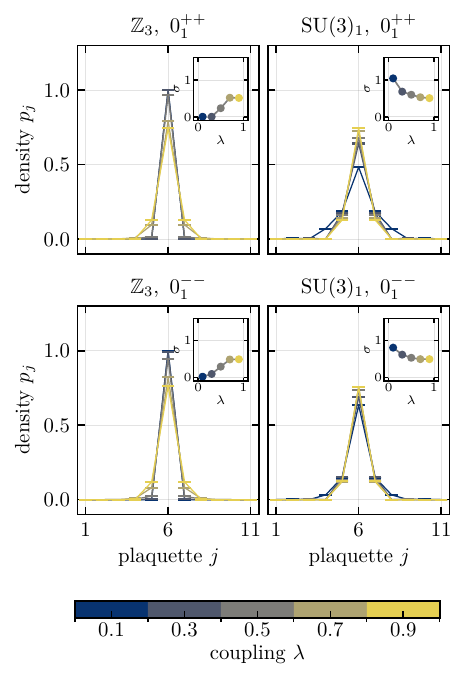}
    \caption{Probability distribution $p_j$ derived from the energy density excess (as in Eq. \eqref{eq:wannierprobability}) of the MLWFs for different quasiparticle bands and couplings $\lambda$. 
    The main plots show the spatial distribution of the energy excess, while the insets show the standard deviation $\sigma$ (the spread) of the energy-density distribution for each case.}
    \label{fig:wanniers}
\end{figure}

In \cref{fig:wanniers}, we present the MLWFs for both the $\Z{3}$ and $\SU{3}_1$ models. 
In the strong coupling limit (high $g^2$, large $\lambda$), the flat bands imply that the MLWFs are supported on a single site, reflecting the local nature of the excitations. 
As the coupling $\lambda$ is reduced, the differences between the Abelian and non-Abelian cases become apparent. 
In the Abelian case, the quasiparticles remain highly localizable even as $\lambda$ decreases; the spread 
$\sigma$ remains small and approaches zero, indicating that the MLWF remains concentrated on a single site (or a very small number of sites). 
In contrast, in the non-Abelian case, reducing $\lambda$ leads to a significant increase in the spread $\sigma$: the MLWFs spread over several sites, preventing the formation of strictly local quasiparticles in the weak coupling regime.

\begin{figure}[t]
    \centering
    \hspace*{-5mm}\includegraphics{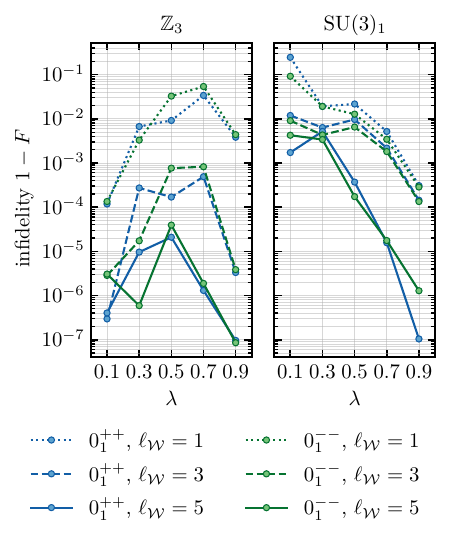}
    \caption{Infidelity $1 - F$ of the variational unitary 
    creation operator ansatz (see \cref{subsec:dressedmpo}) as a function of the coupling $\lambda$, for both $\Z{3}$ and $\SU{3}_1$ across different glueball bands. 
    $\ell_\mathcal{W} = 5$ has been chosen for all the subsequent wave-packet preparation protocols.}
    \label{fig:fidelity}
\end{figure}

In \cref{fig:fidelity}, we report the final infidelity $1-F$ (where $F$ is defined in \cref{eq:procrustes}) of the creation operator after variational optimization, which quantifies the accuracy of the unitary ansatz in reproducing the target MLWF from the vacuum. 
The infidelity follows a similar trend as the MLWF spread shown in \cref{fig:wanniers}: it is small in the Abelian case for all couplings, but increases significantly for $\SU{3}_1$ as $\lambda$ decreases. 
This is expected, since a more delocalized Wannier function requires a more expressive creation operator to be faithfully represented, and the fixed-support unitary ansatz becomes insufficient when the quasiparticle dressing extends over many sites.

Additionally, several Wannier interpolation supports $\ell_\mathcal{W}$ were tested. 
In all cases, an improvement in the fidelity $F$ is observed as the support increases. 
In subsequent simulations, the largest support $\ell_\mathcal{W} = 5$ was used to ensure high wave packet quality.

\subsection{Large system single-quasiparticle spectrum}
\label{subsec:gsingle}

In \cref{subsec:dressedmpo} we introduced the 
construction of the quasiparticle dressed creation 
operators in the form of MPOs, optimized for 
intermediate system sizes $l$. 
However, because a large system of size $L \gg l$ has 
a different vacuum state, it is essential to verify 
that these operators effectively create the intended 
quasiparticle excitations in the larger system. 
To this end, we present a method based on the spectral 
density from the two-point correlator of 
$\hat\phi^\dagger_j$.

\begin{figure}[t]
    \centering
    \hspace*{-0mm}\includegraphics{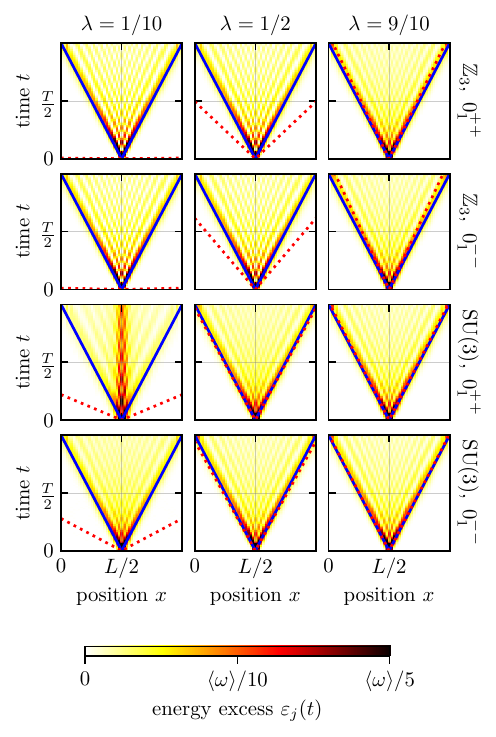}
    \caption{Energy density excess $\varepsilon_j$ (Eq. 
    \eqref{eq:enexcess}) for a time-evolved MLWF 
    $e^{-i\hat{H}t} \hat{\phi}_j^\dagger |\Omega_L\rangle$ on a 
    lattice of $L = 51$ sites for different bands and groups. 
    Energy is shown in units of $0^{++}_1$ band centroid 
    $\langle \omega \rangle$ (\cref{eq:centroid}).
    The solid blue lines indicate the trajectories with 
    maximum speed $v_{\text{max}}$ computed from 
    interpolation of the single-particle spectrum at an 
    intermediate system size 
    ($l=11$, see \cref{fig:disprel}). 
    The dashed red lines represent the maximum speed 
    obtained from second-order perturbation theory in 
    the strong coupling limit (large $g^2$). 
    The total simulation time is $T=L/(2v_{\text{max}})$.}
    \label{fig:cones}
\end{figure}

In \cref{fig:cones}, we show the time-evolution of the 
energy density excess for different groups and 
couplings, quenching an initial state $|\Phi_j(0)\rangle$ 
centered at site $j_0 = L/2$. 
The resulting structure clearly exhibits a ``lightcone'' 
shape, allowing for a direct observation of the maximum 
propagation speed of the quasiparticle. 
Although the simulation is performed on a much 
larger system ($L = 51$) than the one used for the 
spectral analysis, the lightcone edges are in agreement 
with the maximum speed $v_{\text{max}}$ 
(solid blue lines) computed from the interpolation of 
the single-particle dispersion relation at an 
intermediate system size ($l=11$, see \cref{fig:disprel}).
This agreement provides a non-trivial a posteriori validation of the scale-separation conjecture introduced in \cref{subsec:spectral}.
In the strong coupling limit (large $g^2$), the results 
also match the maximum speed predicted by second-order 
perturbation theory (dashed black lines). 
However, as $\lambda$ is decreased ($g^2 \to 0$), the 
perturbative expansion clearly breaks down, while the 
correspondence with the intermediate-size calculations 
remains consistent.

To compute the spectral density, we first define the single-quasiparticle propagator
\begin{equation}
\label{eq:add-correlator}
\Delta_{j - j_0}(t) = \langle \Omega_L | \hat{\phi}_j(t) 
\hat{\phi}_{j_0}^\dagger(0) | \Omega_L \rangle \ ,
\end{equation}
which tracks the amplitude and phase of the quantum process of a quasiparticle initially at $j_0$ being found at $j$ after time $t$. 
We notice that \cref{eq:add-correlator} can be computed as
\begin{equation}
\Delta_{j-j_0}(t) = e^{-i\omega_{\Omega}t} \langle \Phi_{j}(t) 
| \Phi_{j_0}(0) \rangle \ ,
\end{equation}
where $|\Phi_{j}(t)\rangle = e^{-i\hat{H}t} \hat{\phi}_{j}^{\dagger} |\Omega_L\rangle$ is the approximation of the MLWF in the large system, centered at site $j$ and evolved up to time $t$ using the time-evolution algorithm (TDVP).
Here, $\hat{\phi}_{j}^{\dagger}$ is the MPO creation operator and $|\Omega_L\rangle$ is the large DMRG vacuum state.

Since the simulation is performed in a finite system of length $L$ and for a finite time $T$, we actually have access to the propagator \eqref{eq:add-correlator} only in a finite spacetime window:
\begin{equation}
\label{eq:windowed}
\tilde{\Delta }_{j-j_{0}}( t) =\chi _{[ 0,T]}( t) \chi _{[ 1,L]}( j) \Delta _{j-j_{0}}(t) \, 
\end{equation}
where $\chi_{[a,b]}(x)$ is the indicator function which takes values $1$ if $x \in [a,b]$, and $0$ otherwise.
From this simulation data we obtain the propagator in momentum space
\begin{equation}
\label{eq:spectral-density}
\tilde{\Delta }( \omega ,k) = \int_{-\infty}^{+\infty}\frac{dt}{2\pi L}\sum _{j=-\infty }^{+\infty } e^{-i( kj-\omega t)}\tilde{\Delta}_j (t) \, .
\end{equation}
Multiplying a signal by a finite (rectangular) window \eqref{eq:windowed} in the space-time domain corresponds, by the convolution theorem, to convolving its Fourier transform with the window’s transform in momentum space. This produces spectral broadening (e.g., sinc-like side lobes from a rectangular window) but does not shift the position of the peaks in the true spectrum \cite{harrisUseWindowsHarmonic1978}.

\begin{figure}[t]
    \centering
    \includegraphics{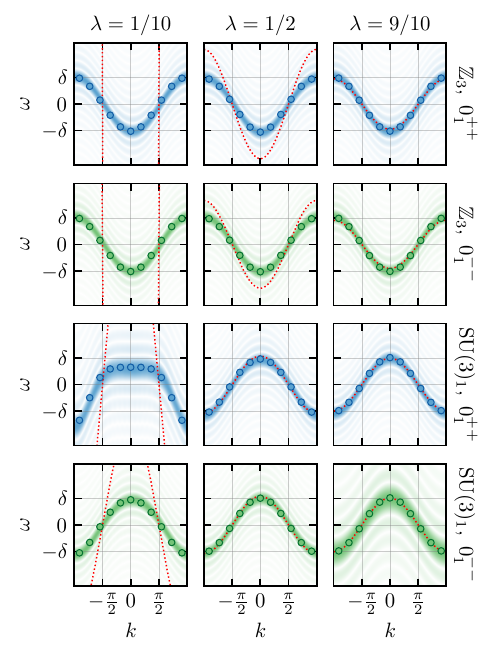}
    \caption{Heatmap of  $|\tilde\Delta(\omega,k)|$ 
    in momentum space (see \cref{eq:spectral-density}) for the 
    lightest glueball bands $0^{++}_1$, $0^{--}_1$ across 
    different couplings $\lambda = 0.1, 0.5, 0.9$ in both 
    the $\Z{3}$ and $\SU{3}_1$ models.
    The densities $\tilde\Delta$ are computed on 
    a lattice of $L = 51$ sites using TDVP-evolved MLWFs 
    (\cref{fig:cones}), and normalized to the maximum value of 
    $\tilde\Delta$ for each model and coupling.
    The spectral-leakage side lobes are due to the space-time 
    window truncation of the time-dependent correlator 
    (see \cref{eq:windowed}).
    The bullets show the first-band eigenvalues $(\omega,k)$ 
    obtained from ED at $l=11$ (\cref{fig:disprel}).
    The color blue ($0^{++}$) and green ($0^{--}$)
    distinguish the two different quasiparticle bands 
    (sectors, see \cref{tab:sectors}).
    The red dashed lines show the dispersion relation 
    obtained from second-order perturbation theory in the 
    strong-coupling limit. In the energy axis, $0$ is the band
    centroid $\langle \omega \rangle$, while $\delta$ is 
    half of the bandwidth.}
    \label{fig:spectralfunc}
\end{figure}

\Cref{fig:spectralfunc} shows the momentum-space propagator \eqref{eq:spectral-density} computed with the windowed real-space propagator \eqref{eq:windowed}. 
The spectral distribution matches the single-particle dispersion relation obtained from intermediate system sizes, confirming that the dressed creation operators introduced in \cref{subsec:dressedmpo} provide a robust ansatz for quasiparticle excitations even in larger systems. 
This method allows verification of the correct propagation of the quasiparticle for all momenta; this procedure represents an efficient self-verification strategy to check that the single-quasiparticle excitation propagates ballistically without decaying into other sectors above the interacting vacuum.

\subsection{Glueball scattering dynamical simulations}
\label{subsec:scattering}

To investigate the dynamical properties of the model, we simulate glueball-glueball scattering processes. 
We prepare an initial state representing two Gaussian wave packets colliding at the center of a large system with $L = 101$ sites. 
Each wave-packet creation operator $\hat{\Psi}^\dagger$ is constructed following the protocol detailed in Appendix \ref{app:wavepacket}, using the dressed creation operator $\hat{\phi}_j^\dagger$ of the scalar glueball $0^{++}_1$, optimized as in \cref{subsec:dressedmpo}.
The wave-packet coefficients are given by
\begin{equation}
c_j = \exp\left[-\frac{(j-j_0)^2}{4\sigma^2} \pm 
ik_0(j-j_0)\right],
\end{equation}
where $j_0$ sets the initial positions and $k_0$ is set 
to be the momentum of 
maximum group velocity $\omega^\prime(k_0)$, which ensures 
also minimum wave-packet dispersion during propagation. 
The state $|\Psi(0)\rangle = \hat{\Psi}_1^\dagger \hat{\Psi}_2^\dagger |\Omega_L\rangle$ 
is then time-evolved using TDVP.

\begin{figure}[t]
    \centering
    \hspace*{-3mm}\includegraphics{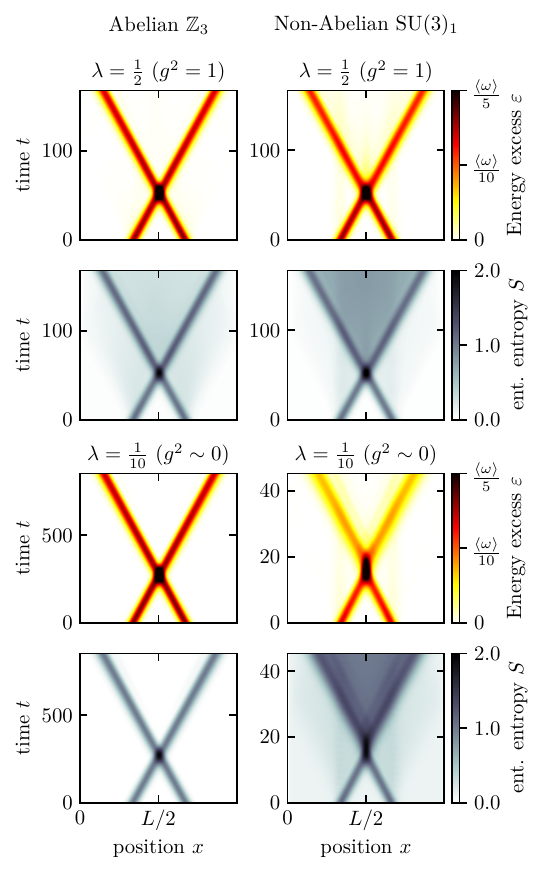}
    \caption{$0^{++}$--$0^{++}$ glueball scattering on a 
    ladder of $L = 101$ plaquettes. 
    On the left column, results for the Abelian $\Z{3}$ model 
    are shown, while on the right column for the non-Abelian 
    $\SU{3}_1$ model. 
    In both cases, we compare $\lambda = 1/2$ (intermediate 
    regime, $g = 1$) and the small-$g$ regime ($\lambda = 0.1$). 
    Local energy excess $\varepsilon_j$ 
    (Eq.~\eqref{eq:enexcess}) and entanglement entropy $S_j$ 
    of the spatial bipartition at site $j$ are shown. 
    Energy is expressed in units of the $0^{++}_1$ band 
    centroid $\langle \omega \rangle$ (see \cref{eq:centroid}). 
    Simulation parameters: wave-packet width $\sigma = L/30$, 
    initial positions $j_0 = L/3, 2L/3$, momentum 
    $k_0 = k(v_{\text{max}})$, maximum bond dimension 
    $\chi = 100$, truncation cutoff $\epsilon = 10^{-10}$, 
    and timestep $dt = T/100$ with $T = L/(2v_{\text{max}})$.}
    \label{fig:scattering}
\end{figure}

The scattering results in \cref{fig:scattering} clearly reveal the fundamental differences between Abelian and non-Abelian theories. 
In the Abelian $\Z{3}$ case, as we approach $g^2 \sim 0$, the quasiparticles propagate almost as free particles, passing through each other with negligible interaction. 
In contrast, the $\SU{3}_1$ case exhibits a clearly interacting nature at $g^2 \to 0$. 
The interaction is most visible in the von Neumann entanglement entropy $S_j(t)$ of the spatial bipartition: for $\SU{3}_1$, the scattering event leads to a significant increase in entanglement at the collision center, whereas for $\Z{3}$ the entropy signal remains close to zero 
near $g^2 \to 0$. 
This interacting behavior in $\SU{3}$ is a direct consequence of the non-Abelian plaquette interactions, which persist even in the weak-coupling limit. 
Since the system is effectively one-dimensional, the only entangling degree of freedom can be a phase shift \cite{Milsted2022CollisionsFalseVacuumBubble}.

\subsection{Glueball detection}
\label{subsec:glueballdetection}

Dressed Wannier creation operators provide a powerful tool to detect single-quasiparticle density during time-evolution. 
By employing the detectors $\hat{\rho}^\phi$ introduced in \cref{subsec:detection} (see \cref{eq:rhodetector}), we can locally identify the presence of specific quasiparticle flavors during and after the collision. 
In this section, we apply this detection protocol to $\SU{3}_1$ scattering simulations in the small-coupling regime ($\lambda = 0.1$, $g^2 \sim 0$), considering scalar ($0^{++}$), pseudoscalar ($0^{--}$), and mixed initial states.
We also detect a lower bound on the resonance energy density $\varepsilon_j[\hat{X}]_{\text{min}}$ as described in Appendix \ref{app:resonances}.

\begin{figure*}
\centering
\includegraphics{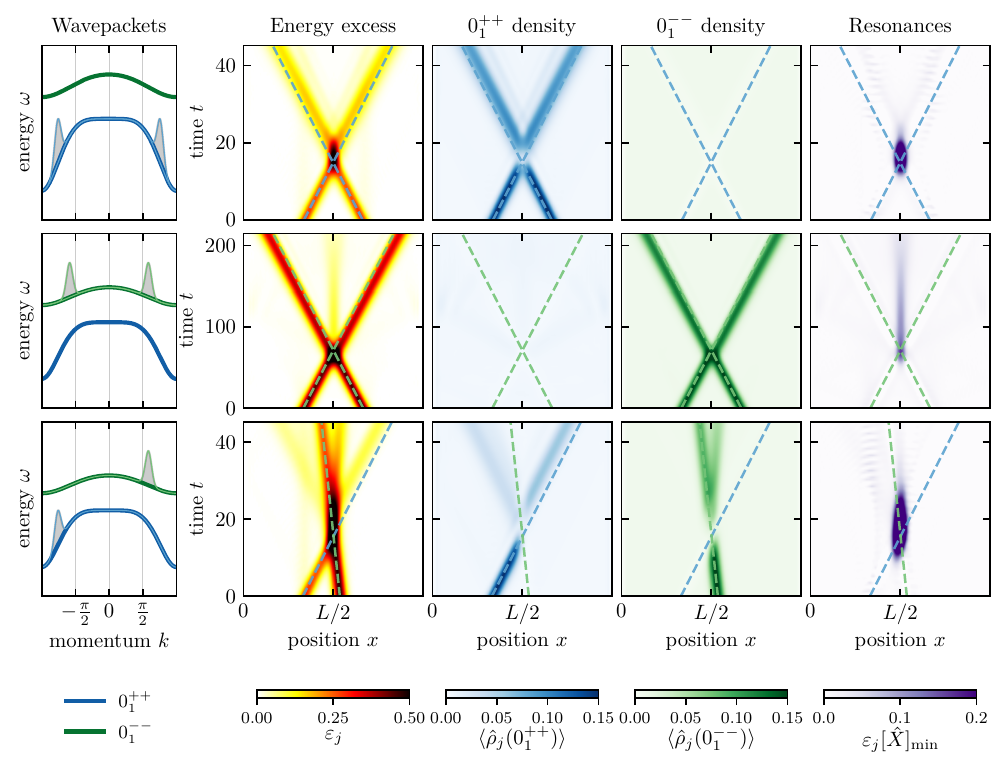}
\caption{
Scattering simulations in the weak coupling regime: $\SU{3}_1$ with $\lambda = 0.1$ ($g^2=1/3$). 
\textbf{Rows:} scalar--scalar glueball ($0^{++}$--$0^{++}$) scattering (first), pseudoscalar--pseudoscalar ($0^{--}$--$0^{--}$) scattering (second), and mixed scalar--pseudoscalar ($0^{++}$--$0^{--}$) scattering (third). 
\textbf{Columns:} (1) Initial single-quasiparticle wave-packet probability density in momentum space; (2) local excess energy density $\varepsilon_j$ (Eq.~\eqref{eq:enexcess}); (3) and (4) species-resolved detection of the $0^{++}$ 
(scalar) and $0^{--}$ (pseudoscalar) glueballs, via the 
projector-like detectors $\hat{\rho}^\phi$ defined in 
\cref{subsec:glueballdetection} 
(Eq.~\eqref{eq:rhodetector}); (5) lower bound on the 
resonance energy density 
$\varepsilon_j[\hat{X}]_{\text{min}}$ 
(see Appendix \ref{app:resonances}). 
The straight lines in the panels are obtained by interpolation of the single-quasiparticle density before the collision, and they represent the expected path of free-propagating quasiparticles. 
Simulation parameters as in \cref{fig:scattering}.}
\label{fig:detect}
\end{figure*}

As shown in \cref{fig:detect}, the detection signals reveal interesting dynamical behavior. 
We first examine scalar--scalar scattering ($0^{++}$--$0^{++}$), where a short-lived resonance $X$ is formed:
\begin{equation}
\label{eq:channels-scalar}
0^{++} 0^{++} \rightarrow X \rightarrow 0^{++} 0^{++}.
\end{equation}
In the center of the collision, the detection signal for $0^{++}$ vanishes, revealing the presence of a two-$0^{++}$ quasiparticle state or a new resonant state(s) $X$. 
No significant formation of $0^{--}$  is observed in this channel.

For pseudoscalar--pseudoscalar scattering ($0^{--}$--$0^{--}$) we observe contributions from 
two channels: a non-interacting channel, and a channel with formation of a long-lived resonance $Y$:
\begin{subequations}
\label{eq:channels-pseudo}
\begin{align}
0^{--} 0^{--} &\rightarrow 0^{--} 0^{--}, \label{subeq:pseudo-pass} \\
0^{--} 0^{--} &\rightarrow Y \rightarrow \, \, ? \, \, , \label{subeq:pseudo-res}
\end{align}
\end{subequations}

Finally, we consider the mixed scattering between scalar 
and pseudoscalar glueballs ($0^{++}$--$0^{--}$), which 
exhibits both a transmission channel, where quasiparticles 
maintain their original direction, and a (lower amplitude) reflection 
channel, where they swap:
\begin{subequations}
\label{eq:channels-mixed}
\begin{align}
0^{++} 0^{--} &\rightarrow Z \rightarrow 0^{++} 0^{--}, 
\label{subeq:mixed-trans} \\
0^{++} 0^{--} &\rightarrow Z^\prime \rightarrow 0^{--} 0^{++}. 
\label{subeq:mixed-ref}
\end{align}
\end{subequations}
Understanding the nature of these resonances $X, Y, Z^{(\prime)}$ is beyond the scope of the present work.
What can be stated with confidence is that the resonance bands intersect the multi-quasiparticle continuum bands in the spectrum of the large system.

The results presented here demonstrate the ability of the Wannier-based detection protocol to resolve scattering channels. 
While the current focus is on understanding the physical dynamics, this framework could be extended to extract quantitative observables such as matrix elements, phase shifts, and time delays. 
We emphasize that our approach provides direct access to real-time dynamics, offering insights that go beyond what is typically available from static non-perturbative techniques.

\section{Conclusions and Outlook}
\label{sec:conclusions}

We introduced a model-independent framework to prepare and detect localized quasiparticle excitations on top of strongly correlated many-body ground states.
We demonstrated our method by simulating real-time scattering events in pure $\Z{3}$ and (truncated) $\SU{3}$ LGTs on a ladder.
In the $\SU{3}$ case, we observed a rich scattering phenomenology already within the minimal hardcore-gluon truncation \cite{Cataldi20232+1DSU2YangMills, Rigobello2023Hadrons1plus1DHamiltonian}, to be contrasted with mostly trivial collisions for $\Z{3}$.

The studied $\SU{3}$ gauge model is mappable to a qutrit chain: the shown mapping absorbs gauge constraints into the effective qutrit DoFs, avoiding explicit symmetry handling in the numerical implementation. This enables a direct implementation of the model on qutrit-based quantum-simulation platforms \cite{Calajo2024DigitalQuantumSimulation}. 

Our variational scheme optimizes unitary 
operators rather than states, distinguishing it from the 
tangent-space quasiparticle ansatz, making the method 
more amenable to digital quantum simulation platforms. 
To this goal, the resulting wave-packet creation operators may be decomposed into sequences of Givens rotations 
\cite{Chai2025FermionicWavePacket,Chai2025ScalableQuantumAlgorithm}, or synthesized via quantum optimal control protocols \cite{omranGenerationManipulationSchrodinger2019},  providing a practical route toward model-independent wave-packet state preparation on near-term quantum hardware.
The framework further allows for the detection of 
quasiparticle excitations via local measurements, 
enabling the resolution and tracking of excitations 
in collision experiments on quantum simulators.

While the discussion here was limited to one spatial 
dimension, MLWFs are well defined in any dimension 
\cite{Marzari2012MaximallyLocalizedWannier}. 
Extensions to higher dimensions are therefore in principle 
possible, albeit limited by the moderate system sizes 
attainable by ED or Krylov methods, upon which the 
extraction of dispersion relations relies. However, this 
limitation can be overcome, as methods already exist to 
extract excitation spectra of finite systems directly 
within a tensor-network (TN) ansatz 
\cite{vandammeEfficientMatrixProduct2021a}.
Moreover, the incorporation of topological excitations into the algorithm can, in principle, be achieved, representing a promising direction for future work (e.g., via twisted boundary conditions in intermediate system sizes).
Although the current localization procedure assumes 
isolated bands, generalizations to (same sector) crossing 
single-quasiparticle bands would allow tackling more 
complex spectra; notably, such formalisms are already well 
established in the Wannier localization literature 
\cite{Marzari2012MaximallyLocalizedWannier}.
Finally, future benchmarks should explore matter coupled 
LGTs, such as one-dimensional $\SU{2}$ 
\cite{Cataldi20232+1DSU2YangMills} and two-flavor 
QCD \cite{Rigobello2023Hadrons1plus1DHamiltonian}.

\section{Acknowledgments}

We warmly thank G. Cataldi, M. Di Liberto, G. Guarda, J. C. Halimeh, and L. Maffi for fruitful private discussions.

The research leading to these results
has received funding from the following organizations:
the European Union
via NextGenerationEU Project No. CN00000013$-$Italian Research
Center on HPC, Big Data and Quantum Computing
(ICSC),
via the Quantum Technology Flagship project PASQuanS2.1 
and the Horizon 2020 Research and Innovation Programme 
under the Marie Skłodowska-Curie 
Grant Agreement No. 101034267;
the Italian Ministry of University and Research (MUR)
via PRIN2022 project TANQU,
and via Departments of Excellence 
2023-2027 project Quantum Frontiers;
the World Class Research Infrastructure - Quantum
Computing and Simulation Center (QCSC) of Padova
University;
The Italian Istituto Nazionale di Fisica
Nucleare (INFN) via Iniziativa Specifica project IS-Quantum;
Regione Veneto via
program PR Veneto FESR 2021-2027 project CONVECS.

\appendix

\section{Wannier functions}

\label{app:wannierfunctions}

The assumption of translational invariance in periodic 
systems allows one to compute Bloch states $\{| k\rangle \}$, 
which form an orthonormal basis of the single-quasiparticle 
Hilbert space. To maximally localize these states 
in real space, one can naively compute the discrete 
Fourier transform
\begin{equation}
| \phi _{j} \rangle =\frac{1}{\sqrt{l}}\sum _{k} e^{ikj}| 
k\rangle \ ,\ \label{eq:wannierdef}
\end{equation}
where $j$ labels the lattice sites and $l$ is the number 
of unit cells. This transformation is unitary, hence the 
set $\{| \phi _{j} \rangle \}$ is orthonormal whenever the 
Bloch states are orthonormal. However, each Bloch state is 
identified only up to a global phase, $| k\rangle 
\rightarrow e^{i\varphi _{k}}| k\rangle $. Consequently, 
the Fourier transform is not uniquely defined but depends 
on the choice of these phases
\begin{equation}
| \phi _{\theta } \rangle =\frac{1}{\sqrt{l}}\sum _{k} 
e^{ikj+i\varphi _{k}}| k\rangle =\frac{1}{\sqrt{l}}
\sum _{k} e^{i\theta _{k}}| k\rangle 
\ ,
\end{equation}
where, in the last step, we absorbed $ 
kj+\varphi _{k} =\theta _{k}$, since $ 
\varphi _{k}$ is arbitrary. Given a set $ 
\{\theta _{k}\}$, the orthonormal basis $ 
\{| \phi _{\theta } \rangle \}_{j}$ is called Wannier 
functions (WFs) in the gauge $ \theta $. This 
gauge freedom can be exploited to impose desirable 
properties on $ | \phi _{\theta } \rangle $, 
most notably maximal localization in real space 
around a given cell. The problem of Wannier localization 
is thus equivalent to selecting a set $\{\theta _{k}\}$ 
over the Brillouin zone that minimizes the spatial spread 
of $| \phi _{\theta } \rangle $.

Since, for a generic interacting model, the single-quasiparticle 
density is not well-defined, our approach consists of localizing 
the quasiparticle by exploiting the locality of the interaction, 
as described in \cref{subsec:wannierloc}. Starting 
from the energy excess \eqref{eq:enexcess}, we can compute 
$ \langle \Omega |\hat{h}_{j} |\Omega \rangle $ 
by exploiting the translational invariance of the vacuum 
$ \langle \Omega |\hat{h}_{j} |\Omega \rangle =\langle \Omega |\hat{h}_{j+1} |\Omega \rangle $, which 
gives:
\begin{equation}
\langle \Omega |\hat{h}_{j} |\Omega \rangle =\frac{1}{l}
\sum _{j} \langle \Omega |\hat{h}_{j} |\Omega \rangle 
=\frac{1}{l} \langle \Omega |\hat{H} |\Omega \rangle 
=\frac{\omega_{\Omega}}{l} \ .
\end{equation}
By translational invariance and using the definition 
\eqref{eq:wannierdef}, it follows that
\begin{align*}
\langle \phi _{\theta } |\hat{h}_{j} |\phi _{\theta } 
\rangle  & =\frac{1}{l}\sum _{k,k^\prime } e^{i( 
    \theta _{k} -\theta _{k^\prime })} \langle k^\prime 
    |\hat{h}_{j}| k\rangle \\
 & =\frac{1}{l}\sum _{k,k^\prime } e^{i( \theta _{k} 
 -\theta _{k^\prime })} \langle k^\prime |\hat{T}^{j}
 \hat{h}_{0}\hat{T}^{\dagger j}| k\rangle \\
 & =\frac{1}{l}\sum _{k,k^\prime } e^{i( \theta _{k} 
 -\theta _{k^\prime } +jk^\prime -jk)} \langle k^\prime 
 |\hat{h}_{0}| k\rangle \ ,
\end{align*}
where, without loss of generality, we set $j = 0$ as the 
central site.
The quantities $ \langle k^\prime 
|\hat{h}_{0}| k\rangle $ can be stored in an 
$ l\times l$ matrix, which can be computed 
efficiently before the variational minimization process.

If the Hamiltonian has inversion symmetry $\hat{P}$, 
with the origin chosen at an inversion center ($j = 0$), 
a MLWF can be chosen as an inversion eigenstate, 
and inversion-breaking gauges cannot yield a 
strictly smaller minimum global spread 
\cite{Marzari2012MaximallyLocalizedWannier}.
This allows us to choose a gauge such that
\begin{equation}
\hat{P} |\phi _{\theta } \rangle = \pm |\phi _{\theta } 
\rangle \ .\ \label{eq:symcondition}
\end{equation}
Here, $\pm$ denotes the parity of the 
$k=0,\pi$ states. Substituting \eqref{eq:wannierdef} into 
\eqref{eq:symcondition}, applying 
$\hat{P}|k\rangle = \pm |-k\rangle$, and exploiting the 
independence of $|k\rangle$, it is straightforward to derive 
that
\begin{equation}
\sum _{k} e^{i\theta _{-k}}| -k\rangle =\sum _{k} 
e^{i\theta _{k}}| k\rangle \ .\ \label{eq:symcondition2}
\end{equation}
Multiplying both sides by $\langle k^\prime |$ 
in \eqref{eq:symcondition2}, we obtain
\begin{equation}
\theta _{k} =\theta _{-k} \quad \bmod 2\pi \ .
\label{eq:symgauge}
\end{equation}
The gauge \eqref{eq:symgauge} halves the number of 
variational parameters and also implies symmetry in 
the energy excess $ \varepsilon _{-j} 
=\varepsilon _{j}$. The minimization can then proceed 
with the functional $ \sigma ^{2}[ \theta ]$, 
as described in \cref{subsec:wannierloc}.

Marzari and Vanderbilt systematically extended the 
minimization algorithm to multi-band subspaces, allowing 
the construction of well-localized Wannier functions even 
in the presence of band degeneracies 
\cite{Marzari1997MaximallyLocalizedGeneralized,
Marzari2012MaximallyLocalizedWannier}. 
Future work is required to extend and test this 
energy-based spatial localization to degenerate bands and 
higher dimensions.

\section{Wavepacket state preparation and detection}

\label{app:wavepacket}

In this section, we describe in detail the procedure for 
constructing the creation operator of the Wannier function 
from the Wannier state, as well as for constructing the 
creation operator of a generic wave packet.

\subsection{Variational problem for the creation operator}

Given the vacuum state $|\Omega \rangle$ and a MLWF state 
$|\phi_{j} \rangle$ centered at site $j$, we want to find an operator $\hat{\phi}_{j}^{\dagger}$ such that 
$\hat{\phi}_{j}^{\dagger} |\Omega \rangle \simeq |\phi_{j} 
\rangle$ with maximum fidelity. This operator must be (a) localized in the 
Wannier support $\mathcal{W}$ of length $\ell_\mathcal{W}$ (this assumption enables the construction of 
wave-packets) and (b) unitary, so that it preserves the norm of 
the state:
\begin{equation}
\hat{\phi}_{j}^{\dagger} = \hat{\phi}_{\mathcal{W}}^\dagger 
\otimes \mathbb{I}_{\overline{\mathcal{W}}} \, , \qquad \hat{\phi}_{\mathcal{W}}^\dagger \hat{\phi}_{\mathcal{W}} = \mathbb I \, .
\label{eq:phioperator}
\end{equation}
The choice of unitarity imposes a stronger condition than 
mere norm preservation, enabling subsequent gate 
decomposition and implementation in Givens-like rotations 
\cite{Chai2025FermionicWavePacket,
Chai2025ScalableQuantumAlgorithm}.
In this way, $\hat\phi_j^\dagger$ can be 
applied to wave-packet preparation protocols in quantum 
simulation platforms.

The variational problem imposes to maximize the fidelity
\begin{equation}
F[\hat{\phi}_{\mathcal{W}}^\dagger] = | \langle 
    \phi_j | (\hat{\phi}_{\mathcal{W}}^\dagger \otimes 
    \mathbb{I}_{\overline{\mathcal{W}}}) |\Omega \rangle 
    |^2 \, . \label{eq:fidelity}
\end{equation}
keeping $\hat\phi_\mathcal{W}^\dagger$ unitary.
This is a generalization to unitary matrices and 
limited support of the known orthogonal Procrustes 
problem \cite{schonemannGeneralizedSolutionOrthogonal1966}, 
whose solution is obtained as follows.

Let us define the operator acting on the Wannier support:
\begin{equation}
\hat{A}_{\mathcal{W}}^\dagger \equiv \Tr_{\overline{\mathcal{W}}} 
|\phi_{j} \rangle \langle \Omega |. \label{eq:Aoperator}
\end{equation}
In order to solve the optimization problem is convenient 
to write the fidelity in Eq.~\eqref{eq:fidelity} as
\begin{equation}
F[\hat{\phi}_{\mathcal{W}}^\dagger] = \left| 
    \Tr_{\mathcal{W}}[(\Tr_{\overline{\mathcal{W}}} 
    |\Omega \rangle \langle \phi_j |\right)
    \hat{\phi}_{\mathcal{W}}^{\dagger}]|^2 
    = | \Tr_{\mathcal{W}}[\hat{A}_{\mathcal{W}}
    \hat{\phi}_{\mathcal{W}}^{\dagger}]|^2\ .
\end{equation}
Performing an SVD, we obtain
\begin{equation}
\hat{\phi}_{\mathcal{W}}^{\dagger} = \hat{U}_{\phi} 
\hat{\Sigma}_{\phi} \hat{V}_{\phi}^{\dagger}, \quad 
\hat{A}_{\mathcal{W}}^\dagger = \hat{U}_{A} \hat{\Sigma}_{A} \hat{V}_{A}^{\dagger},
\end{equation}
where $\hat \Sigma_\phi = \mathbb{I}$ because we imposed $\hat\phi^\dagger_\mathcal{W}$ to be unitary. From the von Neumann trace inequality 
\cite{hornMatrixAnalysis1985}, it follows that
\begin{equation}
\label{eq:fidelitybound}
F[\hat{\phi}_{\mathcal{W}}^{\dagger}] \leqslant 
|\Tr[\hat\Sigma_{A} \hat\Sigma_{\phi}]|^2 = |\Tr[\hat\Sigma_{A}]|^2 = \Vert 
\hat{A}_{\mathcal{W}}^\dagger \Vert_{1}^2 \, .
\end{equation}
This inequality is saturated (a solution of the variational problem) if and only if 
$\hat{\phi}_{\mathcal{W}}^\dagger$ and $\hat{A}_{\mathcal{W}}^\dagger$ 
share all the singular vectors, i.e.,
\begin{equation}
\hat U_{\phi} = \hat U_{A}, \quad \hat V_{\phi} = \hat V_{A}\ ,
\end{equation}
which leads to the optimal solution
\begin{equation}
\hat{\phi}_{\mathcal{W}}^{\dagger} = \hat U_{A} \hat V_{A}^{\dagger} \, . 
\label{eq:optimal_phi}
\end{equation}
All other solutions are related by $\hat{\phi}_{\mathcal{W}}^{\dagger} \mapsto \hat{\Lambda}
\hat{\phi}_{\mathcal{W}}^{\dagger}\hat{\Lambda}^{\dagger}$, 
where $\hat{\Lambda}$ acts trivially in the operator support of 
$\hat{A}_{\mathcal{W}}^{\dagger}$, and as a generic 
unitary transformation in the null-space of 
$\hat{A}_{\mathcal{W}}^\dagger$ (numerically, where singular values $\hat\Sigma_{A}$ are below a threshold $\epsilon$).

\subsection{Wave packets from the creation operator}

Once the dressed creation operator 
$\hat{\phi}_{j}^{\dagger}$ is obtained, a generic 
single-quasiparticle wave packet can be constructed as
\begin{equation}
\label{eq:wavepacket_operator}
\hat{\Psi}^{\dagger} = \frac{1}{\mathcal{N}}\sum_{j=1}^{L} 
c_{j} \hat{\phi}_{j}^{\dagger}, 
\end{equation}
where $c_{j}$ are generic complex coefficients and 
$\mathcal{N} = \sqrt{\sum_{j} |c_{j}|^{2}}$ is the 
normalization factor. The wave-packet state is then 
obtained by applying this operator to the vacuum of the 
large system:
\begin{equation}
|\Psi \rangle = \hat{\Psi}^{\dagger} |\Omega \rangle. 
\label{eq:wavepacket_state}
\end{equation}

The coefficients $c_{j}$ can be chosen to prepare 
wave packets with desired properties, such as specific 
center-of-mass position, momentum, and width. For example, 
a Gaussian wave packet centered at position $j_{0}$ with 
momentum $k_{0}$ and width $\sigma$ can be prepared by 
choosing
\begin{equation}
c_{j} = \exp\left[-\frac{(j-j_{0})^{2}}{4\sigma^{2}} + 
ik_{0}(j-j_{0})\right]. \label{eq:gaussian_wavepacket}
\end{equation}
Since the dressed creation operators 
$\hat{\phi}_{j}^{\dagger}$ are represented as MPOs, the 
wave-packet operator $\hat{\Psi}^{\dagger}$ can be 
efficiently constructed in the MPO representation by a 
linear combination of MPOs. The application to the vacuum 
state can then be performed efficiently using standard 
MPS-MPO contraction algorithms 
\cite{schollwockDensitymatrixRenormalizationGroup2011,
fishmanITensorSoftwareLibrary2021}.

To perform the MPO summations of \cref{eq:wavepacket_operator}, one can mainly adopt two strategies:
\begin{enumerate}
\item start from $\hat \Psi_1^\dagger = c_1 \hat \phi_1^\dagger$ and perform iterations of summations $\hat \Psi_{n}^\dagger = \hat \Psi_{n-1}^\dagger + c_n \hat \phi_n^\dagger$ with MPO compressions after each summation, till the final result $\hat \Psi^\dagger = \hat \Psi_L^\dagger$;
\item if the compressions of the previous approach are too resource-expensive, a more efficient approach is to construct the automata picture of the MPO from the creation operator, as in \cite{Rigobello2020ScatteringProcessesTensor} or employing specific TN summation algorithms \cite{Pavesic2025ScatteringInducedFalse}.
\end{enumerate}
In this work, the first approach has been implemented, since the controlled bond dimension allowed the compression without heavy computational time.

\subsection{Detection of resonances}
\label{app:resonances}

In this section, we describe the procedure to detect the presence of resonances and unknown quasiparticles in scattering simulations with a TN approach.
To this aim, we introduce a nonlinear functional, designed to suppress contributions 
associated with known quasiparticles while isolating the 
residual contribution due to uncharacterized excitations. 
Although this procedure does not admit an interpretation 
in terms of a physical measurement, it proves practically (see \cref{subsec:glueballdetection}) to be a robust 
numerical filtering scheme for identifying 
many-body resonances.

Let us consider two known quasiparticle reduced density 
matrices $\hat{\rho}_{1}, \hat{\rho}_{2}$, as defined in 
\cref{subsec:detection}. We start by 
defining the following operator
\begin{equation}
\hat{X} = \mathbb{I} - c_{1}\hat{\rho}_{1} - 
c_{2}\hat{\rho}_{2}, \label{eq:X_operator}
\end{equation}
where $c_{1}$ and $c_{2}$ are positive real coefficients. 
Defining the non-linear functional for the operator $\hat O$,
\begin{equation}
\varepsilon_{j}[\hat{O}] = |\langle \psi(t) 
|\hat{O}\hat{h}_{j} |\psi(t) \rangle - \langle \Omega 
|\hat{O}\hat{h}_{j} |\Omega \rangle |, 
\label{eq:nonlinear_functional}
\end{equation}
which reduces to the excess density in 
Eq.~\eqref{eq:enexcess} for $\hat{O} = \mathbb{I}$, it 
follows from the triangular inequality
\begin{equation}
\varepsilon_{j}[\hat{X}](t) \geqslant \varepsilon_{j}(t) 
- c_{1} \varepsilon_{j}[\hat{\rho}_{1}](t) - c_{2} 
\varepsilon_{j}[\hat{\rho}_{2}](t) \equiv 
\varepsilon_{j}[\hat{X}]_{\text{min}}(t). 
\label{eq:triangular_inequality}
\end{equation}
The initial state lies entirely in the subspace spanned 
by the ``free'' sectors described by $\hat \rho_{1}$ and 
$\hat \rho_{2}$. There is no component orthogonal to their 
support that contributes to the excess energy. Under this 
assumption, $\varepsilon_{j}[\hat{X}](0) = 0$ is consistent 
at $t = 0$. Hence, for an initial scattering state, we 
require $\varepsilon_{j}[\hat{X}]_{\text{min}}(0) = 0$ 
otherwise, a contradiction follows 
$\varepsilon_{j}[\hat{X}](0) \geqslant 0$. Thus, $c_{1}$ 
and $c_{2}$ can be tuned to have 
$\varepsilon_{j}[\hat{X}]_{\text{min}}(0)$ vanish for all $j$. 
Once these coefficients are obtained by interpolation, 
Eq.~\eqref{eq:triangular_inequality} provides a lower 
bound on the resonance contribution to the energy.

This can be straightforwardly generalized to an 
arbitrary number of known particles identified by 
$\hat{\rho}_{\alpha}$:
\begin{equation}
\hat{X} = \mathbb{I} - \sum_{\alpha} 
c_{\alpha}\hat{\rho}_{\alpha}, \label{eq:X_general}
\end{equation}
with the corresponding bound
\begin{equation}
\varepsilon_{j}[\hat{X}](t) \geqslant 
\varepsilon_{j}(t) - \sum_{\alpha} c_{\alpha} 
\varepsilon_{j}[\hat{\rho}_{\alpha}](t). 
\label{eq:general_bound}
\end{equation}
Note that having 
$\varepsilon_{j}[\hat{X}]_{\text{min}}(t) > 0$ 
for $t > 0$ is a sufficient condition for the 
presence of resonances or uncharacterized excitations 
in the scattering products. The spatial profile of 
$\varepsilon_{j}[\hat{X}](t)$ provides information about 
the localization of these resonant contributions, enabling 
their identification and characterization during the 
scattering dynamics.

\section{Tensor Network methods}
\label{app:tnmethods}

To study the low-lying spectrum and out-of-equilibrium 
dynamics of quantum many-body (QMB) systems, tensor 
network techniques can be employed.
These methods are based on variational ansatz and sparse 
representations of quantum many-body states that exploit 
their entanglement structure, with MPS \cite{Fannes1992FinitelyCorrelatedStates} being the 
most prominent example.
A QMB state is defined on a lattice $\Lambda$ with 
$N = |\Lambda|$ sites. 
Each lattice site $i$ is associated with a local Hilbert 
space $\mathcal{H}_i$ of dimension $d_{i}$.
The state $\ket{\psi}$ of the quantum system therefore 
resides in the composite Hilbert space
\begin{math}
    \mathcal{H} = \bigotimes_{i \in \Lambda} 
    \mathcal{H}_{i}
    \ .
\end{math}
The quantum state $\ket{\psi}$ can then be expanded as
\begin{equation}
    \ket{\psi} = \sum_{i_{1}, \ldots, i_{N}} \psi_{i_{1}, 
    \ldots, i_{N}} \ket{i_{1}, \ldots, i_{N}}
    \ ,
\end{equation}
where $\{\ket{i}_{j}\}_{i=1}^{d_{j}}$ is an orthonormal 
basis for the local Hilbert space $\mathcal{H}_{j}$.
A complete description of the quantum state $\ket{\psi}$ 
therefore requires specifying $\prod_{j}^{N}d_{j}$ complex 
coefficients, which becomes computationally intractable 
for large $N$.
However, many physically relevant quantum many-body states, 
such as low-energy eigenstates and certain states 
encountered during out-of-equilibrium dynamics, exhibit 
only moderate entanglement.
For low-energy eigenstates, this behavior is characterized 
by the entanglement area law 
\cite{Hastings2007AreaLawOnedimensional}, which states 
that the entanglement entropy of a subsystem typically 
scales with its boundary area rather than its volume 
\cite{Eisert2010ColloquiumAreaLaws}.
When the area law holds, the quantum state $\ket{\psi}$ 
admits an efficient representation in terms of a 
MPS \cite{Eisert2010ColloquiumAreaLaws}
\begin{equation}
    \ket{\psi} = \sum_{i_1,\dots,i_{N}}
    \Tr\left[A_{1}^{i_{1}} \cdots A_{N}^{i_{N}}\right] 
    \ket{i_{1},\dots,i_{N}},
    \label{eq: MPS}
\end{equation}
where the number of parameters scales polynomially with 
the system size. In~\cref{eq: MPS}, for each physical 
index $i_{j}$, the matrix $A_{j}^{i_{j}}$ has shape 
$\chi_{j-1} \times \chi_{j}$, where $\chi$ is the bond dimension. In the case of open boundary 
conditions, the boundary matrices reduce to row and column 
vectors, such that $\chi_{1} = \chi_{N} = 1$. In one-dimensional 
systems with gapped local Hamiltonians, where locality 
refers to interactions between neighboring sites, the area 
law has been proven analytically \cite{Hastings2007AreaLawOnedimensional}. 
In this work, we employ the DMRG algorithm 
\cite{Schollwock2005DensitymatrixRenormalizationGroup} 
to compute an estimate of the ground state.
To simulate time evolution within the TN
framework, TDVP is employed 
\cite{Haegeman2011TimedependentVariationalPrinciple}, 
which does not rely on a Suzuki–Trotter decomposition. 
In this approach, the dynamics are constrained to the 
chosen TN manifold, here, the MPS manifold with fixed bond 
dimension $\chi$, by projecting the Hamiltonian onto the 
tangent space and solving the time-dependent Schr\"odinger 
equation within that subspace.

\section{Hamiltonian formulation of LGT}
\label{app:hamiltonian}

We consider a square lattice $\Lambda$ with a set of sites 
$\{s\}$ hosting matter DoFs connected by directed 
links $\{\ell\}$ hosting gauge DoFs. We also identify 
the set $\{\square\}$ of minimal loops (square plaquettes). 
We consider the pure gauge theory, 
i.e., only gauge DoFs remain in the infinite 
mass limit of matter, allowing the latter to be 
integrated out.

Let $G$ be a compact Lie group, which we take to be the gauge group. 
In this work we focus on two cases: the Abelian $G = \U{1}$ and the non-Abelian $G = \SU{3}$, representing QED and QCD, respectively. 
In the Hamiltonian formulation of gauge theories, the Hilbert space of a gauge DoF on a link $\ell$ is $\mathcal{H}_{\ell} = L^2(G)$, spanned by the group basis $\{|g\rangle_{\ell}\}_{g \in G}$ \cite{Zohar2015FormulationLatticeGauge}, such that the total Hilbert space of the pure theory is $\mathcal{H} = \bigotimes_{\ell} \mathcal{H}_{\ell}$.

The group acts via the generator of gauge transformations 
$\hat{G}_s$; all (physical) gauge-invariant states 
$|\psi\rangle \in \mathcal{H}$ satisfy the Gauss' law 
constraint \cite{Zohar2015FormulationLatticeGauge}. For a compact Lie group $G$ this constraint can be written in terms of generators of gauge transformations:
\begin{equation}
\hat{G}_s |\psi\rangle = 0 \quad \forall\, s \in 
\Lambda.
\end{equation}
More generally, lattice gauge theories can also be defined for discrete or finite groups (e.g., $\Z{3}$), where infinitesimal generators do not exist; in that case Gauss’ law is imposed by requiring physical states to be invariant under the action of the local gauge transformation operators associated with each site.

Given an irrep $j$ of $G$, the parallel transporter $\hat{U}_{mn}^j$ is an operator-valued matrix whose entries act on the group basis as $\hat{U}_{mn}^j |g\rangle = D_{mn}^j(g) |g\rangle$, where $D^j(g)$ are the unitary representation matrices of $j$ \cite{Zohar2015FormulationLatticeGauge}.

The plaquette operator is the Wilson loop around the plaquette
\begin{equation}
\hat{U}_{\square} = \Tr\, \mathcal{P} \prod_{\ell \in \square} \hat{U}_{\ell},
\end{equation}
where $\mathcal{P}$ denotes path-ordering around the loop and the trace is over internal group indices. 
To define the electric energy operator $\hat{E}_{\ell}^2$, a convenient basis for $\mathcal{H}_{\ell}$ is the irrep basis $\{|jmn\rangle_{\ell}\}$ \cite{Zohar2015FormulationLatticeGauge}, in which $\hat{E}_{\ell}^2$ acts diagonally as
\begin{equation}
\hat{E}^2 |jmn\rangle = C_2(j) |jmn\rangle,
\end{equation}
where $C_2(j)$ is the quadratic Casimir of the irrep $j$.

The Kogut--Susskind Hamiltonian of the pure theory \cite{Kogut1975HamiltonianFormulationWilsons,Zohar2015FormulationLatticeGauge,milstedQuantumYangMillsTheory2018} reads
\begin{equation}
\hat{H}_{\text{KS}} = \frac{g^2}{2}\hat{H}_E 
+ \frac{1}{2g^2}\hat{H}_B \label{eq:kshamiltonian}
\end{equation}
with electric and magnetic terms
\begin{equation}
\hat{H}_E = \sum_{\ell} \hat{E}_{\ell}^2, \qquad 
\hat{H}_B = - \sum_{\square} \hat{U}_{\square} 
+ \hat{U}_{\square}^{\dagger},
\end{equation}
and $[\hat{H}_{\text{KS}}, \hat{G}_s] = 0$ at every site.

For better control of $g^2 \in [0,\infty)$, we rescale 
the Hamiltonian and introduce a different reparameterization of the coupling:
\begin{equation}
\hat{H} \to \frac{2\lambda}{g^2}\hat{H}, \qquad 
\lambda = \frac{g^4}{1+g^4},
\end{equation}
so that $\lambda = 0$ for $g = 0$ and $\lambda = 1$ for 
$g^2 \to \infty$. The Hamiltonian \eqref{eq:kshamiltonian} 
can then be written as
\begin{equation}
\hat{H} = \lambda \hat{H}_E + (1-\lambda)\hat{H}_B.
\end{equation}
The Hilbert space dimension of $\mathcal{H}_{\ell}$ is 
infinite for compact Lie groups $G$, so numerical 
simulation requires a finite-dimensional truncation. 
Common strategies include quantum link models 
\cite{Chandrasekharan1997QuantumLinkModels}, 
$q$-deformations of Lie algebras 
\cite{zacheQuantumClassicalSpinNetwork2023}, 
and irrep truncation (used in this work; see 
\cref{sec:model}).

\section{Hamiltonian matrix elements}
\label{app:rishonoperators}

\subsection{Dressed site formalism}
\label{app:dressed_site_formalism}

\begin{figure*}[t]
  \centering
  \input{images/junction_basis}
  \caption{The $\top$-junction basis for a single vertex, generated by assigning all allowed truncated link states and subsequently projecting onto the gauge-invariant singlet state.
    In this notation, $r,g,b$ represent the basis states of the fundamental representation $\boldsymbol{3}$ of $\SU{3}$, while the states $c,y,m$ are the basis states of the anti-fundamental representation $\bar{\boldsymbol{3}}$. 
    The state $w$ is the unique state of the singlet representation.}
  \label{fig:tjunctionbasis}
\end{figure*}

\begin{figure*}[t]
  \centering
  \input{images/plaq_mels}
  \caption{Non-vanishing expectation values of the corner operators for the $\top$-junction states listed in \cref{fig:tjunctionbasis}. 
  Using these operators, one can compute the plaquette operator $\hat{U}_\square$, whose 27 non-vanishing matrix elements are also represented in this figure.}
  \label{fig:plaquettemels}
\end{figure*}

In LGTs, by convention, matter DoFs are located on lattice sites, while gauge DoFs reside on the links.
For numerical simulations with continuous gauge groups, the infinite-dimensional link Hilbert space associated with the gauge fields must be truncated.
To implement the truncation of gauge DoFs, we work in the irrep basis, where the quadratic Casimir operator is diagonal.
The irrep basis is related to the group-element basis, in which the magnetic energy is diagonal, through the Peter--Weyl theorem \cite{Peter1927VollstandigkeitPrimitivenDarstellungen}.
In this basis, states $\ket{j\,m\,n}$ are labeled by the irrep $j$ and the associated internal indices $m$ and $n$.
The bosonic Hilbert space can then be truncated by restricting the allowed eigenvalues of the quadratic Casimir operator.
In the present work, we employ the lowest nontrivial truncation, retaining the trivial irrep and the states in the fundamental and anti-fundamental irreps, both of which have the Casimir eigenvalue $4/3$.

To locally enforce Gauss's law in LGT, we employ the dressed-site formalism \cite{Silvi2014LatticeGaugeTensor}.
This approach reduces the original non-Abelian constraint to an Abelian one.
The resulting dressed sites typically have larger local Hilbert-space dimensions than those arising from other truncation schemes, but can still be efficiently treated within TN simulations.

First, the parallel transporter is decomposed into the left and right DoFs, denoted by the left (L) and right (R) half-link operators.
This is achieved by embedding the link Hilbert space as
$\ket{j m n} \;\to\; \ket{j m}_L \otimes \ket{j n}_R$.
The parallel transporter then takes the form
\begin{equation}
  \hat{U} \to \sum_{k}
  \zeta^{L(k)\alpha \, \dagger}_{x,+\mu}\zeta^{R(k)\beta}_{x+\mu,-\mu} \,
  \label{eq_from_U_to_rishons}
  \,.
\end{equation}
Physical configurations are those in which the left and right half-links are in the same irrep.
This construction thus yields an Abelian link symmetry on the TN level \cite{ Silvi2014LatticeGaugeTensor}, regardless of whether the gauge symmetry of the theory is Abelian or non-Abelian.
For each matter site $x$, the neighboring half-link DoFs are fused into a single super-site.
Gauss's law then becomes an onsite constraint, identifying the physical (dressed-site) Hilbert space with the super-site states transforming as singlets.
The latter can be expanded in terms of the matter and half-link bases via Clebsch--Gordan decomposition.

\subsection{Electric term operator mapping}
\label{app:electricmapping}

Starting from the expression for the electric energy
\begin{equation}
\label{eq:HE_full}
\hat{H}_{E} = \sum_{j} \hat{E}_{j}^{\uparrow 2}
+ \sum_{j} \hat{E}_{j}^{\downarrow 2}
+ \sum_{j} \hat{E}_{j}^{r\,2} \,,
\end{equation}
we represent the local link DoFs by
\begin{equation}
\label{eq:sigma_link}
\hat{\sigma}_{\ell} \equiv e^{i\hat{n}_{\ell}\varphi}
= \begin{pmatrix}
1 &  & \\
 & e^{i2\pi/3} & \\
 &  & e^{i4\pi/3}
\end{pmatrix} \,.
\end{equation}
By combining \cref{eq:electricfielddiag} with
\cref{eq:sigma_link}, one finds
\begin{equation}
\label{eq:E2_clock}
\frac{\hat{E}_{\ell}^{2}}{C_{2}}
= \begin{pmatrix}
0 &  & \\
 & 1 & \\
 &  & 1
\end{pmatrix}_{\ell}
= \frac{2 - \hat{\sigma}_{\ell}
       - \hat{\sigma}_{\ell}^{\dagger}}{3} \,,
\end{equation}
which gives the expression for the electric field on the legs (in the zero-polarization sector):
\begin{equation}
\label{eq:E2_legs}
\frac{\hat{E}_{j}^{\uparrow 2}}{C_2}
= \frac{\hat{E}_{j}^{\downarrow 2}}{C_2}
= \frac{2 - \hat{\sigma}_{j}^{\uparrow}
       - \hat{\sigma}_{j}^{\uparrow\dagger}}{3} \,.
\end{equation}
For the rungs, we note that the fusion rules modulo~$3$ at a top junction yield
\begin{equation}
\label{eq:sigma_rung}
\hat{\sigma}_{j}^{\dagger}\hat{\sigma}_{j+1}
= e^{i(\hat{n}_{j+1}^{\uparrow}
     - \hat{n}_{j}^{\uparrow})\varphi}
= e^{i\hat{n}_{j}^{r}\varphi}
= \hat{\sigma}_{j}^{r} \,.
\end{equation}
Combining \cref{eq:E2_clock} with \cref{eq:sigma_rung}, we obtain
\begin{equation}
\label{eq:E2_rungs}
\frac{\hat{E}_{j}^{r\,2}}{C_{2}}
= \frac{2 - \hat{\sigma}_{j}^{\uparrow\dagger}
         \hat{\sigma}_{j+1}^{\uparrow}
       - \hat{\sigma}_{j}^{\uparrow}
         \hat{\sigma}_{j+1}^{\uparrow\dagger}}{3} \,.
\end{equation}
Finally, substituting \cref{eq:E2_rungs,eq:E2_legs} into
\cref{eq:HE_full} and identifying
$\hat{\sigma}_{j}^{\uparrow} = \hat{\sigma}_{j}$ as the
chain DoFs, we arrive at
\begin{equation}
\label{eq:HE_potts}
\hat{H}_{E} = -\frac{C_{2}}{3} \sum_{j}
\bigl(\hat{\sigma}_{j}^{\dagger}\hat{\sigma}_{j+1}
+ 2\,\hat{\sigma}_{j} + \text{H.c.}\bigr) \,.
\end{equation}
Note that this identification is exact within the
truncated field space and holds for both
$\Z{3}$ and $\SU{3}_{1}$.
The expression~\eqref{eq:HE_potts} is the interaction
term of a clock model with an additional longitudinal
field, which renders the system non-integrable.

\subsection{Plaquette operator matrix elements}
\label{app:operators}

In this subsection, we outline the computation of matrix 
elements of the parallel transporter in the $\top$-junction 
basis.
The $\top$-junction basis, shown in \cref{fig:tjunctionbasis}, 
comprises all singlet states localized on a lattice site 
with three adjacent links and therefore satisfies Gauss's law.
These states include all possible assignments of irreps on 
the links within a chosen truncation of the Casimir 
eigenvalues, provided that the tensor product of the 
corresponding representations at the junction contains a 
singlet. On the right-hand side of 
\cref{fig:tjunctionbasis}, we also illustrate how the 
singlet is formed.
The label $w$ denotes the unique state in the $\mathbf{1}$ 
representation, while $r,g,b$ label the three states of 
the $\mathbf{3}$ and $c,m,y$ those of the 
$\overline{\mathbf{3}}$.

In the same manner, one could also construct the 
corresponding states for $\U{1}$ as the gauge group.
In this case, however, the construction is trivial, 
since each irrep is one-dimensional, leading to fewer 
possible combinations that yield a singlet upon fusion 
of the attached irreps.

Within this basis, we compute the matrix elements of the 
parallel transporter. This is achieved by first evaluating 
the matrix elements of the corner operator, using the fact 
that the parallel transporter can be expressed as a product 
of corner operators.
The corner operator $\hat{C}$ acts on pairs of adjacent 
links, namely the downward-facing half-link and the link 
to its right.
The operator $\hat{C}$ can be written as a product of 
half-link operators.
If the parallel transporter transforms in the fundamental 
irrep, then the half-link operator acting on the 
downward-facing link is in the fundamental irrep, while 
the half-link operator acting on the rightward link 
transforms in the anti-fundamental irrep.
The matrix elements of the corner operator in the 
$\top$-junction basis are reported in \cref{fig:plaquettemels}.

Using these results, the matrix elements of the plaquette 
operator can be computed.
The resulting matrix elements in the plaquette basis are 
shown in \cref{fig:plaquettemels}.

\bibliography{gauge_scattering_1D}

\end{document}